\documentclass[11pt,a4paper]{article}
\pdfoutput=1
\usepackage[title,titletoc]{appendix}
\usepackage{jheppub}
\usepackage{slashed}
\usepackage[labelformat=empty]{subfig}
\usepackage{graphicx}
\usepackage{caption}
\usepackage{multirow}
 
\title{Finding viable Models in SUSY Parameter Spaces with  Signal Specific Discovery Potential}
\author[a]{Thomas Burgess}
\author[a]{Jan \O ye Lindroos}
\author[a]{Anna Lipniacka}
\author[a]{Heidi Sandaker}
\affiliation[a]{Department of Physics and Technology, University of Bergen, Norway}
\emailAdd{thomas.burgess@ift.uib.no}
\emailAdd{jan.lindroos@ift.uib.no}
\emailAdd{heidi.sandaker@ift.uib.no}
\emailAdd{anna.lipniacka@ift.uib.no}

\abstract{Recent results from ATLAS gives a Higgs
mass of 125.5~\GeV, further constrain already
highly constrained supersymmetric models such as
pMSSM or CMSSM/mSUGRA. Finding potentially discoverable and
non-excluded regions of model parameter space is becoming
increasingly difficult.
Several groups have invested large effort in studying the consequences of Higgs mass bounds, upper limits on rare $B$-meson decays, and limits on relic dark matter density
on  constrained models, aiming at predicting superpartner masses,
and establishing likelihood of SUSY models compared to that of the Standard Model
vis-\'a-vis experimental data. In this
paper a framework for efficient search for
discoverable, non-excluded regions of different SUSY spaces  
giving specific experimental signature of interest is presented. 
The method employs an improved Markov Chain Monte Carlo (MCMC) scheme 
exploiting  an iteratively updated likelihood function to
guide search for viable models. Existing experimental
and theoretical bounds as well as the LHC discovery potential are taken into account. 
This includes recent bounds on relic dark matter density, the Higgs sector and rare
$B$-mesons decays. A clustering algorithm is applied to classify selected models  according to expected phenomenology enabling
automated choice of experimental benchmarks and regions to be used
for optimizing searches. The aim is to provide experimentalist with
a viable tool helping to target experimental signatures to search for, once a class of
models of interest is established. As an example a
search for viable CMSSM models with $\tau$-lepton signatures 
observable with the 2012 LHC data set is presented. 
In the search 105209 unique models were probed. 
From these, ten reference benchmark points covering different ranges of
phenomenological observables at the LHC were selected.  
}
\def\m0{\ensuremath{m_0}}

\def\neutralino{\ensuremath{\tilde{\chi}_1^0}}

\def\pt{\ensuremath{p_\mathrm{T}}}

\def\GeV{\ensuremath{\mathrm{GeV}}}
\def\TeV{\ensuremath{\mathrm{TeV}}}

\newcommand{\python}{\texttt{Python}}

\newcommand{\pythia}{\texttt{Pythia}}
\newcommand{\isajet}{\texttt{ISAJET}}
\newcommand{\isared}{\texttt{isaRED}}
\newcommand{\darksusy}{\texttt{darkSUSY}}

\newcommand{\feynhiggs}{\texttt{FeynHiggs}}
\newcommand{\higgsbounds}{\texttt{HiggsBounds}}

\begin{document}

\setlength{\parskip}{0.5cm}
\setlength{\parindent}{0pt}
\maketitle
\flushbottom

\section{Introduction}
\label{sec:Introduction}

Supersymmetry (SUSY) may alleviate many of the problems
associated with
the Standard Model of particle physics (SM) if the mass of the
superpartners lies close to the~\TeV-scale \cite{Nilles:1983ge},
\cite{Haber:1984rc}.
 Furthermore, it provides a
natural dark matter candidate in the form of the Lightest
Supersymmetric Particle (LSP), if $R$-parity is
conserved~\cite{Martin:1997ns}. However, even the simplest SUSY 
extension of the SM, the so called Minimal Supersymmetric Standard Model
(MSSM), introduces over 100 new free parameters making them very
difficult to experimentally constrain. On the other hand
a large part of the  MSSM parameter space is already ruled out,
as it would lead to unobserved phenomena like 
  non-conservation of lepton numbers, flavour changing neutral
  currents or large CP violation~\cite{Nakamura:2010zzi}. It is therefore common practice  
  to look at constrained models that assume a partial unification of
  parameters at some high energy scale and where the dynamics of the
  high energy theory ensures more viable phenomenologies~\cite{Ellis:2001msa}.   
The minimal SUper GRAvity model (mSUGRA)~\cite{PhysRevLett.49.970} is
an example of such a constrained model where the SUSY parameters are
assumed to unify at the GUT scale into five universal parameters, a
common scalar mass $m_0$, a common gaugino mass $m_{1/2}$, the ratio
between the SUSY Higgs vacuum expectation values $\tan\beta$, a common
trilinear Higgs-sfermion coupling $A_0$, and the sign of the Higgsino
mass parameter $\mu$.  In the ``lighter'' version of it, the  so called
constrained MSSM (CMSSM)~\cite{Drees:1992am,Ellis:1997wva,Lipniacka:2001gb} gravitino mass is not forced
to unify at the same scale as other gaugino masses. In NUHM (Non-Universal
Higgs Masses)~\cite{Baer:2005bu,Baer:2004fu} 
models, masses of Higgs bosons do not unify with 
sfermions to the common $m_0$.       

ATLAS and CMS experimental searches for SUSY
usually present results only in two-dimensional slices of the parameter space
of some simplified model 
assuming fixed values for other
parameters~\cite{Stoye:1457447,Bruneliere:1451244}. 
Due to complicated dependence of 
physical masses and thus experimental signatures on 
all the model parameters, it is easy to leave
specific corners of the model space unexplored in such an approach, leaving out
regions where experimental search may have large discovery potential. 
This has led
several theoretical groups~\cite{Buchmueller:2012hv,Fowlie:2012im}
 to reinterpret experimental searches
in different regions of parameter space with help of simplified
simulators of detector response like DELPHES~\cite{Ovyn:2009tx} or 
PGS~\cite{Conway}. This approach
can be relatively reliable for moderately simple experimental signatures
involving jets and missing transverse energy~($\slashed{E_T}$), but it 
cannot be trusted for more difficult experimental objects like photons or tau leptons. 

MCMC based parameter inference has been
successfully employed to find the most viable region of the full parameter
spaces, based on requirements that the models should be in accordance
with recent  experimental
constraints~\cite{Buchmueller:2012hv,Fowlie:2012im}, including
these on the Higgs boson mass and rare $B$-mesons decays. While such scans 
provide a more complete picture of the still allowed regions of
parameter space they do not consider whether these parameter
space regions are within experimental reach. This poses difficulties for
experimentalists when trying to make direct use of the results.  

In this paper, a MCMC-based framework for determining the 
part of non-excluded model parameter space where a given experimental signature can be observed is presented. By adding a signature specific discoverability parameter to the set of current experimental constraints, the interesting regions of the parameter
space are found. Models from these regions are then partitioned according to 
phenomenology using a clustering algorithm to enable an automatized construction of reference points for optimizing experimental searches. This step distinguishes our approach from existing similar frameworks, for example~\cite{mastercode}.    
The procedure is applicable to a wide range of signatures and
models, and is intended as a tool for experimentalist to extend limits to more interesting regions of parameter space. It is important to note that we do not intend to find the true maximal likelihood regions, as the discoverability measure does not reflect any existing constraint. In order to provide a proof of concept, a concrete example
defining a non-excluded part of  CMSSM parameter space
which  could be  discoverable with $\tau$-leptons in the 
2012 LHC data is outlined.    

The paper is structured as follows: section~\ref{sec:AlgTools}
discusses the publicly available software tools
used to calculate low energy CMSSM observables,
 scan and clustering algorithms, as well as
the specific constraints and phenomenological parameters
used. Section~\ref{sec:mSUGRAScan} describes the results of the scan
and the phenomenological reference points  constructed. Appendix~\ref{sec:appendix} explains the 
details of the algorithm implementation and presents cross-checks of the effects
of experimental constraints with other existing results. In
Section~\ref{sec:Conclusions}, a summary and comments on the procedure
are provided.

\section{Algorithms and Tools} 
\label{sec:AlgTools}

Experimental constraints on dark matter relic density $\Omega h^2$ as
well as on rare processes such as $B_s\rightarrow\mu\mu$ and
$b\rightarrow s+\gamma$ set strict bounds on the parameter space of CMSSM
(see for example \cite{Fowlie:2012im}).
 Furthermore, the Higgs boson mass of
$125.5$~\GeV as measured by ATLAS~\cite{ATLAS-CONF-2013-014} is hard to accomodate in CMSSM, making the
fraction of viable models within current experimental reach extremely
small. This renders simple uniform 
scans highly inefficient. A rough random scan made to explore the
parameter dependence in CMSSM, gave a fraction of $10^{-5}$ models in
accordance with current experimental
constraints. Therefore, more advanced techniques need to be employed to get a representative picture of the  discoverable and non-excluded regions of parameter space in an efficient way.     

The approach used in this paper is to employ a likelihood distribution $P$, that
reflects how well models fit the data and their discovery potential, to perform a guided random walk through
parameter space using Markov Chain Monte Carlo~\cite{gelman2010handbook}. 
This increases the 
 search efficiency as the parameter space is sampled according to the 
distribution $P$ thus  less time is spent sampling 
low likelihood regions. In this work an adaptive MCMC is implemented,
where the likelihood map is based on the compatibility of low energy properties 
of CMSSM models with experimental and theoretical constraints, and discovery potential.
These properties are calculated using several publicly available software tools. 

\subsection{Software Tools} 
\label{sec:Tools}

A series of publicly available software tools is used to calculate the low energy parameters
needed to check experimental constraints on the SUSY
models, and to construct the likelihood map used  in the MCMC scan. 
Parameters are passed between the different tools using the
SLHA-interface~\cite{Allanach:2008qq}. The tools are called in sequence
starting with the least computationally costly, and after
each step 
the likelihood is updated based on the available
parameters. Each component ($i$) 
of the likelihood is constructed to have a maximal value of $P_i$=1 so that 
the likelihood always decreases as the chain progresses. This makes it possible 
to check for rejection after every step in the tool sequence, and enables early
termination of the calculations for a large fraction of low likelihood
models.

In the first step, \isajet\ with \isared~\cite{Paige:2003mg} is used to
run the GUT scale universal parameters down to the electroweak scale, calculate
$\textbf{Br} \left( B_s \rightarrow \mu \mu \right)$, and to check whether the models are 
allowed by several theoretical
constraints, including requirements of a
$\neutralino$ LSP and correct electroweak symmetry. In the next step, 
\feynhiggs~\cite{Heinemeyer:1998yj} and
\higgsbounds~\cite{Bechtle:2008jh} are used to recalculate and check if the model
fulfills experimental
constraints on  the Higgs sector. Afterward, the dark
matter relic density, $\Omega h^2$, and $\textbf{Br}\left(b\rightarrow s + \gamma\right)$ is calculated using \darksusy~\cite{Gondolo:2004sc}, 
which also checks against  experimental constraints on sparticle masses 
from LEP $\Delta\rho$ and Z-width
(see for example \cite{PDBook,Abdallah:2003xe}). Finally, 1000 pp signal events at $\sqrt{s}=8~\TeV$ are
generated using \pythia~\cite{Sjostrand:2007gs} in order to get a
leading order estimate of the SUSY cross-section, $\sigma_\mathrm{LO}$, and to calculate 
the fraction of events ($\textbf{Br}_{\tau},\textbf{Br}_{jet}\ldots$)  containing 
respectively at least one $\tau$, $e$, $\mu$,  
jet with  pseudorapidity in the central part of the
detector, $\left|\eta\right|<2.5$, and sufficiently large momentum in the plane
perpendicular to the beam axis, $p_T>20~\GeV$,
and the average number of these objects per SUSY event ($n_{\tau},n_{jet}\ldots$). 
For each of these objects, the average $p_T$ is calculated for the two with
the  highest transverse momentum. The average  missing transverse energy per event, $\slashed{E_T}$, is also calculated. 

The software tools and their employment are summarized in Table~\ref{tab:Tools}.   
\begin{table}[htbp]
  \caption{Software tools and resulting information employed in this work.
 Average  values from
    \pythia\ are for final states objects  with $|\eta|<2.5$ and $p_T>20 \GeV$.}  
  \label{tab:Tools}
  \begin{center}
    {\footnotesize
      \begin{tabular}{|l|l|}
        \hline
        \textbf{Tool} & \textbf{Information used} \\\hline\hline
        \isajet\ 7.83 \& \isared & SUSY masses,
        $\textbf{Br}\left(B_s\rightarrow\mu\mu\right)$\\\hline 
        \feynhiggs\ 2.9.4 \& \higgsbounds\ 3.8.1 & Higgs
        sector\\\hline  
        \darksusy\ 5.1.1 & $\Omega_\chi h^2$,
        $\textbf{Br}\left(b\rightarrow s+\gamma\right)$\\\hline 
        \pythia\ 8.175 & $\sigma_{\mathrm{LO}}$,$\textbf{Br},\left\langle n,
          p_{T1},p_{T2} \right\rangle$ for
        $\tau,e,\mu,\mathrm{jet}$,$\left\langle
          \slashed{E_T}\right\rangle$ \\\hline  
      \end{tabular}}
  \end{center}
\end{table}

\subsection{Likelihood Map and Experimental Constraints} 
\label{sec:Constraints} 

The likelihood map $P$ used to explore CMSSM parameter space is constructed by
combining a likelihood $P_\mathrm{exp}$ based on experimental and
theoretical constraints with an ad-hoc likelihood related to
the expected number of events with tau leptons, $P_\tau$. Here 
$P_\tau$ is based on the probability of producing observable $\tau$-leptons
with $21/fb$ of the LHC data collected in 2012. $P_\tau$ can be
easily replaced by another likelihood function related to observability
of any signal of interest.  The likelihoods
are normalized so that each individual contribution $P_i$ has a
maximal value~$\max(P_i)=1$. Thus, the full likelihood becomes:

\begin{equation}
  P_\mathrm{tot}=P_\mathrm{exp}\cdot P_\tau ~~ and ~~~~ P_{\mathrm{exp}}=\prod_iP_i \,,
\end{equation}

where  $P_i$  are the likelihoods related to experimental
limits and theoretical constraints. Some of  $P_i$ are
either 0 or 1 as specified in
table~\ref{tab:Pexp}. These include most of theoretical
constraints, 
limits checked internally by the software tools used.
For other experimentally measured quantities Gaussian errors are assumed and the resulting
likelihoods are continuous. Gaussian distributions around the central experimental values are used 
for $\textbf{Br} \left( b \rightarrow s + \gamma \right)$, $\textbf{Br} \left(B_s \rightarrow \mu \mu\right)$, and the Higgs mass, while for the relic density a uniform distribution is chosen with a Gaussian tail above the best observational value. The latter accepts models 
with the relic density lower than the recent Planck result~\cite{Ade:2013lta},
allowing for other unknown sources except of CMSSM neutralinos to contribute to the relic density.
The central values and standard deviations used are $\Omega h^2 = 0.1199 \pm 0.0027$ for the relic density~\cite{Ade:2013lta}, $\textbf{Br}\left( b \rightarrow s+\gamma \right) = \left(3.55 \pm 0.42 \right) \cdot10^{-4}$~\cite{Asner:2010qj} for the charmless $b$-quark decay, with a theoretical uncertainty $\sigma_{th}=\pm0.33\cdot10^{-4}$ \cite{DSmanual:2009}, and $\textbf{Br}\left( B_s \rightarrow \mu\mu \right) = \left(3.2 \pm 1.5 \right)\cdot10^{-9}$~\cite{Aaij:1493302}. For the Higgs mass, the combined ATLAS best fit from the $H\rightarrow\gamma\gamma,4l$ channels is used \cite{ATLAS-CONF-2013-014}, with a theoretical uncertainty $\sigma_{th}=\pm1.5 \GeV$ is assumed \cite{Degrassi:2002fi}, giving $m_{h0}=\left(125.5\pm1.7\right)\ \GeV$. The experimental and theoretical constraints are summarized in table~\ref{tab:Pexp}. The 2011 and 2012 ATLAS and CMS results of direct searches for SUSY in $R$-parity conserving channels are not included in the present work.The reason for it is two-fold. Firstly, the high Higgs mass translates in CMSSM into rather high sparticle masses, on the border of the present direct searches sensitivity. Secondly, our aim is to propose precise regions, where this sensitivity should be checked, and not to exclude them from our scans.  Results obtained by ATLAS and CMS experimenters using dedicated detector response simulations to translate the present limits into other regions of parameter space should be more reliable than ones employing only approximate modelling of detectors response. We thus prefer to use this opportunity to provide tools to experimenters so that they can choose somewhat more interesting regions of SUSY parameter space to present their results.

\begin{table}[htbp]
  \caption{Experimental and theoretical constraints used and the
    associated likelihoods}
  \label{tab:Pexp}
  \begin{center}    
    {\footnotesize
      \begin{tabular}{|l|l|l|}\hline
        \textbf{Constraints}&\textbf{Likelihoods}
        $\mathbf{P_i}$&\textbf{Values}\\\hline\hline
        $\neutralino$ LSP, Correct EWSB, No tachyons \ldots& OK: 1 Not
        OK: 0 &\isajet\ 7.81\\
        Sparticle masses, $\Delta\rho$, Z-width& OK: 1 Not OK: 0
        &\darksusy\ 5.0.5\\
        OK Higgs sector& OK: 1 Not OK: 0 &\higgsbounds\ 3.7.0\\\hline 
        $\textbf{Br}\big(B_s\rightarrow\mu\mu\big)$&$\exp\Big[\frac{(\textbf{Br}_\mathrm{B_s}-\mu_\mathrm{B_s})^2}{-2\sigma_\mathrm{B_s}^2}\Big]$
        &$\big[\mu_\mathrm{B_s},\sigma_\mathrm{B_s}\big]=\big[3.2,1.5\big] \cdot10^{-9}$\\
        $\Omega h^2$&$\exp\Big[\frac{(\Omega
          h^2-\min(\Omega
          h^2,\mu_{\Omega}))^2}{-2\sigma_{\Omega}^2}\Big]$ &
        $\big[\mu_{\Omega},\sigma_{\Omega}\big] =
        \big[1.199,0.027\big]\cdot10^{-1}$\\   
        $\textbf{Br}\big(b\rightarrow
        s+\gamma\big)$ &
        $\exp\Big[\frac{(\text{Br}_\mathrm{bsg}-\mu_\mathrm{bsg})^2}{-2\sigma_\mathrm{bsg}^2}\Big]$  
        & $\big[\mu_\mathrm{bsg},\sigma_\mathrm{bsg}\big]=\big[3.55,0.42\big]\cdot 10^{-4}$\\  
        $m_{h0}$ &
        $\exp\Big[\frac{(m_{h0}-\mu_{h0})^2}{-2\sigma_{h0}^2}\Big]$ 
        & $\big[\mu_{h0},\sigma_{h0}\big]=\big[125.5,1.7\big]$\
        \GeV\\[1ex] \hline
      \end{tabular}
    }
  \end{center} 
\end{table}

The discoverability likelihood $P_{\tau}$ is chosen as a Poissonian  \textit{discoverability} measure constructed from the sum of 
likelihoods for observing a given number of tau events, $N_\tau \ge 1$, given the 
expected number of events containing at least one $\tau$, 
$\left \langle N_{\tau} \right \rangle = \textbf{Br}_{\tau} \cdot \mathcal{L} \cdot \sigma_{LO} $.

\begin{equation}
  P_{\tau} = \sum\limits_{N_{\tau} =
    N_{\tau}^{\min}} P(N_{\tau} | 
  \left\langle N_{\tau} \right\rangle) \ \ \ ,\ \ \ P(N_{\tau} | 
  \left\langle N_{\tau} \right\rangle)=\frac{\left\langle N_{\tau} \right\rangle^{N_{\tau}}\exp\big[-\left\langle N_{\tau} \right\rangle\big]}{N_{\tau}!}\,.
\end{equation}

Leading order SUSY cross-section $\sigma_{LO}$ in $pp$ collisions at 8 TeV  center-of-mass energy, luminosity of $21/fb$ and  
the fraction of events containing at least one $\tau$, $\textbf{Br}_{\tau}$, 
as found from \pythia, are used to calculate $ \left \langle N_{\tau} \right \rangle$ above.

Experimental selection in search of specific signal has broadely speaking two steps. 

The first step ensures that a specific experimental signature characterizing the signal is observed  in the detector. In our case this signature consists of taus, jets and missing transverse energy. In order for this step to be fulfilled one needs to make sure that these experimental objects are within fiducial volume of the detector and have transverse momentum above a given threshold. The number of signal events passing this first step can be predicted with relatively good accuracy without using any sophisticated detector description.

In the second step of the selection, specific cuts in order to reject the backgrounds are performed, and some measure of sensitivity is used in order to optimize bakground rejection while keeping as much of the signal as possible. The number of expected signal events after such a selection can vary orders of magnitude depending on specific strategy chosen.

One example of this is the ATLAS $\tau$ search as presented in \cite{ATLAS-CONF-2013-026}, where two different strategies are considered, one for events where exactly 1 $\tau$  is selected and another for events with 2 or more $\tau$ leptons. This difference in selection strategy  gives large differences  in signal selection efficiency. It is clear that precise choice of
strategy needs to be done with precise tools  using reliable detector simulation, background estimation and cutflow optimization, and this can be done only withing experimental collaborations.

The aim of the scan presented here however is to find interesting regions for $\tau$ searches in which this second step of the selection can be performed, because there is enough events passing the first step.

This is why the constructed measure $P_{\tau}$ is based on the number of $\tau$ events observable in the detector rather then a more precise estimate valid for a specific analysis strategy. The main uncertainty in $P_{\tau}$ comes from neglecting NLO corrections to the cross section and theoretical uncertainties in the LO cross section. The NLO corrections for CMSSM can be relatively large, with k-factors of the order of 3 \cite{GoncalvesNetto:2012yt}.

The LO cross section uncertainty estimate is sensitive to variations in renormalisation/factorization scale, parton distribution function and the strong coupling $\alpha_s$, and can lead to uncertainties of order $100\%$, compared to $\sim20\%$ at NLO \cite{Beenakker:1996ch}. In comparison Pythia MC uncertainties are negligible with relative errors of order $\sigma_{MC}\lesssim 0.1$ for reasonable branching fractions $\textbf{Br}_{\tau}\gtrsim 0.1$. The combined uncertainty in $P_{\tau}$ is also much smaller than the uncertainty associated with experimental selection.

Including NLO calculations for each point increases computational time by a factor of five or more, and given the uncertainties related to experimental selection outlined above, the computational gain outweighs the loss in precision. As our ambition is merely to identify which regions of mSUGRA parameter space are more interesting than others, NLO corrections matter only if they vary a lot across the parameter space. The option for including Prospino NLO calculations \cite{Beenakker:hep-ph9611232} is implemented in the package, and is used to get more precise estimates for the proposed benchmark points. Indeed the corrections are quite comparable across proposed the benchmark points.

The search range in CMSSM parameter space follows the suggestion 
in~\cite{Allanach:2005kz}. The ranges for $m_0$, $m_{1/2}$, $A_0$ and $\tan\beta$ are presented in table~\ref{tab:mSUGRArange}. The anomalous muon magnetic moment, $\delta a_{\mu}=a^{\text{exp}}_{\mu}-a^{\text{SM}}_{\mu}$, is not taken into account in this scan, other than as a reason for choosing sign$(\mu)>0$, which is required to give positive SUSY contributions. This is because the value of $\delta a_{\mu}$ is generally incompatible with other constraints \cite{Fowlie:2012im}, leading to maximal likelihood regions in agreement with neither. In addition the actual value of $\delta a_{\mu}$ seems to be open for debate due to uncertainties in both LO and NLO hadronic contributions to $a^{\text{SM}}_{\mu}$ \cite{Fowlie:2012im,Bodenstein:2013flq}
 
 
A fixed value for $m_{\text{top}}=173$~GeV was taken. 
\begin{table}[htbp]
  \caption{Search ranges used for the CMSSM MCMC scans.}
  \label{tab:mSUGRArange}
  \begin{center}
    {\footnotesize
      \begin{tabular}{|l|c|}\hline
        \textbf{Parameter} & \textbf{Range}\\\hline\hline 
        $m_0$ & [60,3000]\ \GeV\\
        $m_{1/2}$ & [60,3000]\ \GeV\\
        $A_0$ & [-5000,5000]\ \GeV\\
        $\tan\beta$ & [2,60] \\
        sign($\mu$) & +1\\
        \hline
      \end{tabular}
    }
  \end{center} 
\end{table}

The lower bounds on the universal masses and on $\tan\beta$ in 
table~\ref{tab:mSUGRArange} stem from LEP~\cite{Schael:2006cr,Abdallah:2003xe} bounds,
while the upper limits are chosen on the basis of
naturalness for the masses $m_0$ and $m_{1/2}$ and perturbativity of
the Yukawa couplings for $\tan\beta$. The range for the trilinear
coupling $A_0$ is extended compared to~\cite{Allanach:2005kz},
since the Higgs mass has a quadratic dependence on
$A_0$~\cite{Niessen:2008hz} allowing for higher $m_{h0}$ at large
values of $|A_0|$. 

It is important to note that the more realistic the likelihood, 
the more computationally efficient is the search for
interesting models. However, the set of models found does not depend on 
fine details of the likelihood function used, as
we accept points in a range of likelihood. Our goal is to find a set of 
interesting models fulfilling latest experimental constraints, in order
to guide further experimental searches. We do not intend to make any 
{\it statistically quantified decision} on which of the selected models
are more likely than others.
 
\subsection{MCMC Algorithm} 
\label{sec:MCMCAlg}

The MCMC method used here is a Metropolis-Hastings algorithm~\cite{1970} where, 
given a point $\mathbf{x}=\left\{x_1,x_2\ldots x_D\right\}$ in a D-dimensional parameter space, a proposal
distribution, $Q(\mathbf{y}|\mathbf{x})$,  is used to sample a new point 
$\mathbf{y}$. The proposal distribution
is related to the likelihood $P$ of  the  new point $\mathbf{y}$ being 
``interesting'' from the point
of view of requirements  described in  section \ref{sec:Constraints}.  
The new
point is accepted randomly with a probability given by 

\begin{equation} \alpha(\mathbf{y}|\mathbf{x})=\min\left(1,\frac{P(\mathbf{y})Q(\mathbf{x}|\mathbf{y})}{P(\mathbf{x})Q(\mathbf{y}|\mathbf{x})}\right)\,,
\label{eq:alpha}
\end{equation}
 
\begin{figure}[htb]
  \begin{center}
  \makebox[\textwidth]{
  \begin{tabular}{cc}
  \subfloat[a) A proposal distribution $Q(\mathbf{y}|\mathbf{x})$ used to sample new points ($\mathbf{y_1},\mathbf{y_2}$) from a point $\mathbf{x}$. These points are either accepted ($\mathbf{y_2}$) or rejected ($\mathbf{y_1}$) depending on the ratio between the underlying likelihood $P$ and $Q$ as given by $\alpha$ \eqref{eq:alpha}]{\label{fig:MCMCstepA}
\includegraphics[width=75mm]{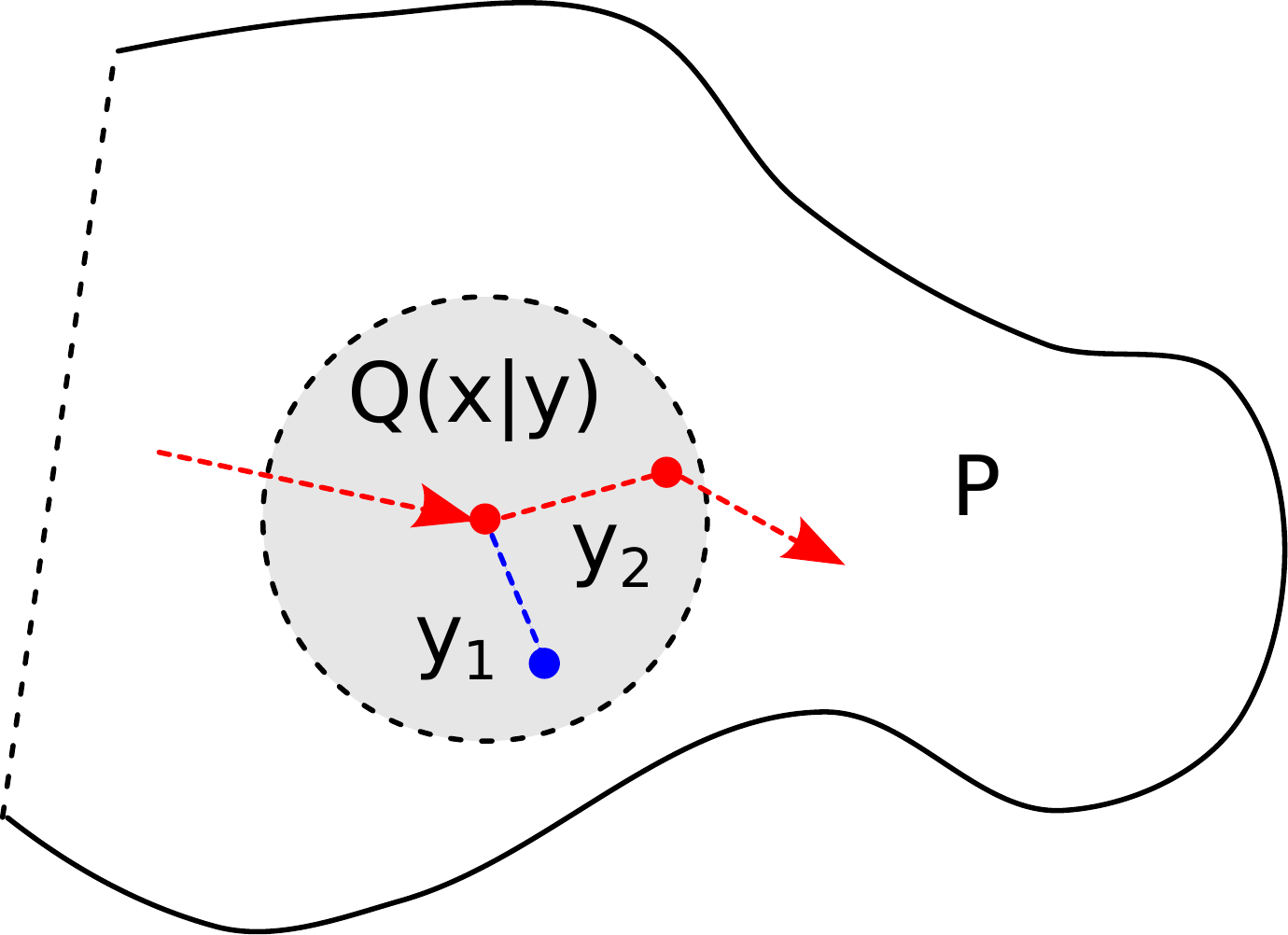}} 
  &
  \subfloat[b) Example MCMC random walk finding narrow high likelihood regions ($B$), when starting from a point in a low ($A$) or zero ($C$) likelihood region.
The grid of dots is shown to  illustrate how such regions can be missed by uniform grid-based scans.]{\label{fig:MCMCstepB}
    \includegraphics[width=75mm]{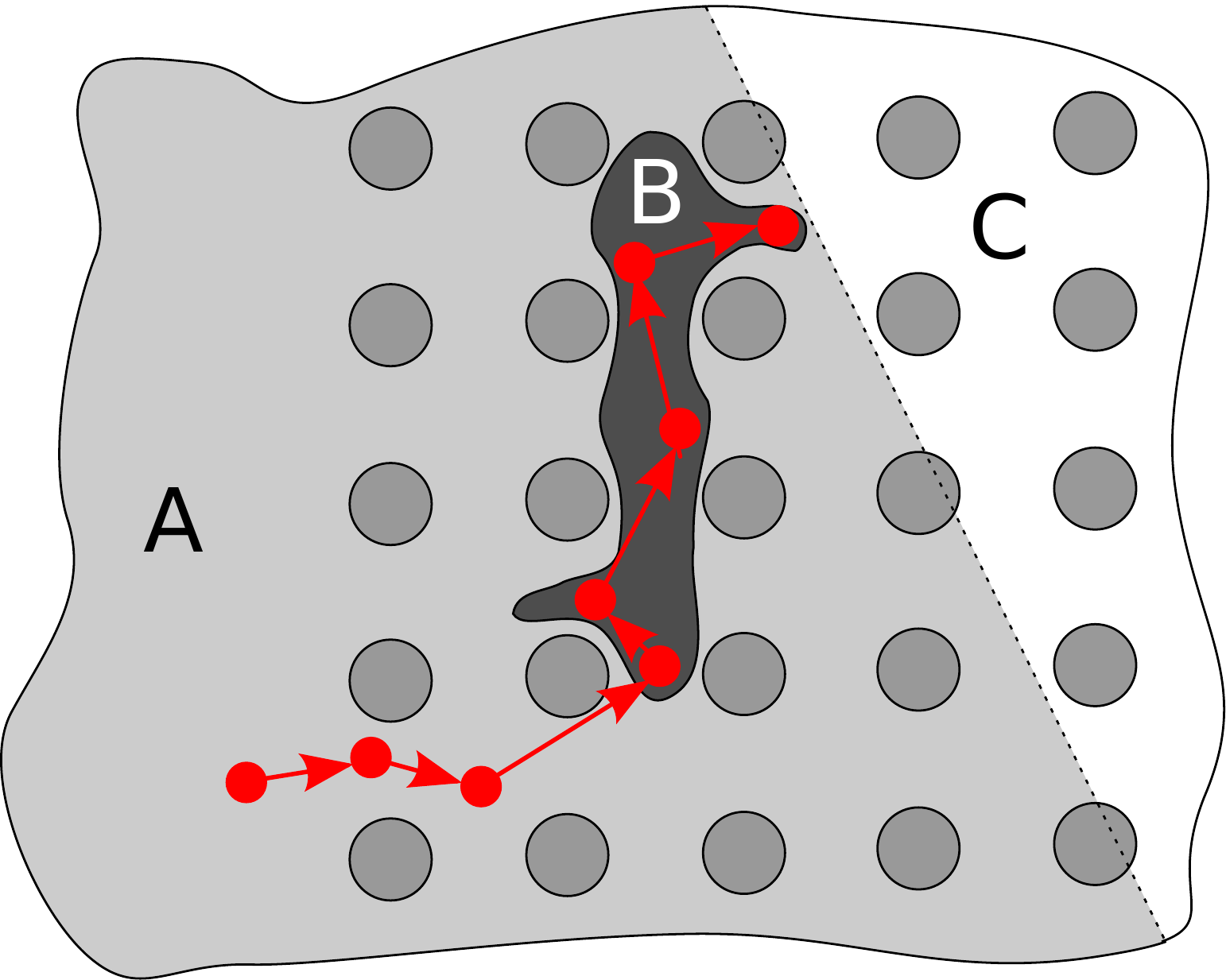}} 
  \end{tabular}}    
  \end{center}
  \caption{Illustrations of the standard MCMC sampling method,~\ref{fig:MCMCstepA}, and a typical MCMC random walk,~\ref{fig:MCMCstepB}} 
  \label{fig:MCMC} 
\end{figure}

If $\mathbf{y}$ is accepted, it is added to the chain and the next point is
sampled starting from $\mathbf{y}$. If it is not accepted,
the chain remains at $\mathbf{x}$ and the process is repeated as 
illustrated in  figure~\ref{fig:MCMC}.
The asymptotic distribution of likelihoods calculated for  the resulting chain of
points is the desired likelihood distribution $P$. 

In order to efficiently map possible high likelihood regions of the parameter space which are separated
by large regions of low likelihood 
a regional adaptive MCMC algorithm similar to~\cite{2009arXiv0903.5292C} has been implemented. 
The algorithm approximates the target likelihood distribution $P$ as a mixture of normalized  multivariate
Gaussian distributions  and uses this approximation as a 
basis for a proposal $Q(y|x)$, as explained in the following section~\ref{sec:AdaptiveMultichainMetropolis}. 
This proposal is used to guide multiple MCMC search chains in parallel and it is iteratively updated according
to the resulting selected sample of points.  

\subsubsection{Adaptive Multi Chain Monte Carlo} 
\label{sec:AdaptiveMultichainMetropolis}
 
An initial estimate for the proposal was constructed by uniformly
sampling the space such that all separated regions where the likelihood  
$P$ is high are covered. The idea is similar to the bank sampling introduced in \cite{Allanach:2007qj}, where prior knowledge about the local maxima of the likelihood distribution is incorporated into the proposal to increase efficiency of sampling the 
distributions where these maxima are separated by large regions of low likelihood.
The points of CMSSM parameter space chosen for the initial sample 
were required to pass all discrete cuts
and  to give experimentally measured physical variables within a 
reasonable range of the experimentally preferred values,
see table~\ref{tab:Pexp} for details.  
Sampled points in CMSSM parameter space were weighted according to their likelihood and
clustered using \emph{$k$-means} algorithm, to be defined in~\ref{sec:Clustering}. 
The "shape" of each  cluster was estimated by  calculating  the weighted mean $\boldsymbol{\mu}$ vector  and covariance matrix $\mathbf{\Sigma}$.
The number of clusters corresponded to the  number of normalized Gaussian distributions (normal mixture)
that was to be used to approximate the likelihood distribution~$P$, as explained below. 

A small fraction of large jumps~\cite{Guan:2006:MCM:1132243.1132248,2009arXiv0903.5292C,Allanach:2005kz} was added to the standard small jumps illustrated in figure~\ref{fig:MCMCstepA} in order
to increase sampling efficiency.  To achieve this  
a global proposal term $q_G(\mathbf{y})$, was added to the standard
local one, $q_L(\mathbf{y}|\mathbf{x})$, giving the full proposal distribution:

\begin{equation}
  Q(\mathbf{y}|\mathbf{x})=\beta q_L(\mathbf{y}|\mathbf{x})+(1-\beta)q_G(\mathbf{y})\,,
\end{equation}

Here  $\beta$ is a  mixing parameter relating  the global and the local
proposal terms, explained further.

The global proposal $q_G(\mathbf{y})$ was taken as 
a set of $m$ multivariate normal distributions,  $\mathcal{N}$,  as in~\cite{west1993}.
The set was  large enough
to describe the main features of the target likelihood. 
Each multivariate distribution was multiplied
by a weight factor $w_i$,
defined in the
formula below.
 
\begin{align}
	q_G(\mathbf{y})=\sum_{i=1}^{m} w_i\mathcal{N}(\mathbf{y}|\boldsymbol{\mu}_i,\mathbf{\Sigma}_{G,i})& \ \ \ ,\ \ \ \mathcal{N}(\mathbf{y}|\boldsymbol{\mu},\mathbf{\Sigma})=\frac{\exp\big[-\frac{1}{2}(\mathbf{y}-\boldsymbol{\mu})^{\mathrm{T}}\mathbf{\Sigma}^{-1}(\mathbf{y}-\boldsymbol{\mu})\big]}{\sqrt{(2\pi)^{D}\left|\mathbf{\Sigma}\right|^{}}}\,,\\
	&w_i=\frac{\sum_{\mathbf{x}_i} P(\mathbf{x}_i)}{\sum_{\mathbf{x}} P(\mathbf{x})}\notag
\end{align}

Each weight factor  was estimated by summing up
the total likelihood over CMSSM parameter space points in a  cluster $i$. 
Here $\boldsymbol{\mu}_i$ is the vector of the means of the $i$'th normal distribution, while
$\mathbf{\Sigma}_{G,i}$ is the covariance matrix of the $i$'th component of the mixture. 
The local proposal $q_L(\mathbf{y}|\mathbf{x})$ was taken as a normal distribution with mean, 
$\mathbf{x}$, and the covariance, $\mathbf{\Sigma}_{L,i}$, characterizing  the closest cluster in the parameter space. 
An euclidean distance measure was used and each parameter was scaled so that the search ranges 
defined in table~\ref{tab:mSUGRArange} varied  from 0 to 1.
The local proposal covariance was chosen so that:
$  \mathbf{\Sigma}_{L,i}=\alpha_i\mathbf{\Sigma}_{G,i} \ ,\  i\in\left\{1,2,\ldots m\right\} $,
where $\alpha_i$ was a parameter 
adapted such that the local acceptance rate for points  in the parameter space region 
within the cluster $i$ was between
  $0.05$  and $0.15$. The rather low acceptance rate was chosen because the hierarchical nature of 
the likelihood calculation yields higher computational speed for low acceptance rates.
Thus our optimal acceptance rate is probably lower than that of $0.23$ found in~\cite{And94weakconvergence}. 
The acceptance probability for stepping from a given point $x$ to a
new point $y$ was then given as:
   
\begin{equation}
  \label{eq:StepProb} 
  \alpha(\mathbf{y}|\mathbf{x})=\min\left(1,\frac{P(\mathbf{y})\left[\beta
        q_L(\mathbf{x}|\mathbf{y})+(1-\beta)q_G(\mathbf{x})\right]}{P(\mathbf{x})\left[\beta
        q_L(\mathbf{y}|\mathbf{x})+(1-\beta)q_G(\mathbf{y})\right]}\right)\,.
\end{equation}

The search chains were started from random CMSSM parameter space  points in the
weighted sample and followed independently. After a given number of steps
data were re-clustered and the proposals were updated, taking into account the new sampled parameter space points, where the new points were weighted according to the estimated likelihood. 
A certain 
likelihood threshold $P_{\min}$  was required for the first relevant point in each
chain, since we are interested in high likehood regions. 
The implementation details are described in
the Appendix~\ref{sec:appendix}.  

\subsection{Clustering Algorithm} \label{sec:Clustering}

A modified \emph{$k$-means}~\cite{0214.46201} algorithm, to be defined
below,  has been devised in 
order to cluster
likelihood-weighted points.
The role of clustering is two-fold. Firstly clusters in CMSSM parameter space were needed to calculate the approximate
Gaussian distributions used in the proposal described in
section~\ref{sec:AdaptiveMultichainMetropolis}. 
Secondly,
sets of high-likelihood model-points in CMSSM parameter space
were clustered according to the different
experimental signatures they were  expected to exhibit in the detectors
at the LHC. 

The \emph{ $k$-means} algorithm defines clusters in the parameter space  by
assigning each point to  the closest centroid,~($C$).
The algorithm was initialized by choosing at random   $k$ points
in the parameter space as cluster centers,~$C$. 
Next, each $C_i$ was refined as the average of the points near 
to it and points  were reassigned to the new $C$, and the procedure 
was repeated until
it converged to a set of stable $C$s.
To increase the speed of the algorithm, a maximum number of iterations
and a minimum improvement between iterations was set for the centroids
positions  refinement.
 
In order to define a closest centroid, 
a distance measure is required. 
An euclidean distance measure in CMSSM space was employed, scaling 
each parameter such that the search ranges 
defined in table~\ref{tab:mSUGRArange} varied  from 0 to 1. Using another distance measure (for example a log scale) only alters the proposal distribution and thus only affects the efficiency of the algorithm, not the results. 
The {\emph{$k$-means}} algorithm, described above, with predetermined number of clusters
was used to cluster points in CMSSM space. 


In order to define reference points in phenomenological space the arbitrary choice of the
number of such points has to be avoided.
To this end \emph{$k$-means} formed the basis for 
a \emph{$g$-means} algorithm, with which the number of clusters 
was determined automatically, as explained further.
In both instances the random  first guess of cluster 
centers (centroids)  positions was improved as suggested in~\cite{ilprints778}, 
($k$-means$++$ algorithm). In this method probability of
picking a new point as a cluster center 
was weighted by the square distance to the
closest already picked point. This guess reduced as well the
average number of iterations to achieve convergence.



To determine the number of clusters $k$ in the space of phenomenological
observables, 
a \emph{$g$-means}-algorithm~\cite{Hamerly03learningthe} was employed.
 It started with 
applying  \emph{$k$-means} for
$k=2$, thus dividing the parameter space into two
sub-clusters.  Then a statistical test to verify  if 
the likelihood distributions
of both  clusters  could be described by a 
single Gaussian was performed. For each phenomenological observable, $x_i$,
the standard deviation, $\sigma_i$,
and the mean, $\bar{x_i}$, were calculated 
and each  observable transformed $x_i'=(x_i-\bar{x_{i}})/\sigma_i$
to facilitate defining the distance for clustering purposes.
  If the test rejected the
single Gaussian distribution hypothesis, the procedure was repeated recursively
for each of the new clusters, otherwise recursion was terminated. The
statistical test was  performed by means of the one dimensional Anderson-Darling
normality measure~\cite{Stephens_1974}. The distances
between  the points and a plane that separated clusters and was perpendicular
to the vector between the two centroids were subjected to the measure above. 
As the clustering could be sensitive both to outliers
and to the random positioning of initial centroids, the whole splitting 
procedure was iterated $n_\mathrm{avg}$ times and the average number 
of clusters $\left\langle k\right\rangle$ was noted. Next, the clustering 
outcome
with $k$ closest to $\left\langle k\right\rangle$ was picked. If
there were several clustering
outcomes giving the same number of clusters $k$, one of these was picked at
random.
In the final step the obtained centroids were subjected to the  \emph{$k$-means}
clustering one more time to ensure that a stable configuration has been
found. 


\section{CMSSM with $\tau$ Signatures}
\label{sec:mSUGRAScan}

\subsection{Results}
\label{sec:Results}

\begin{figure}[htbp]
 \begin{center}
   \makebox[\textwidth]{
   \begin{tabular}{ccc}
   \subfloat{\includegraphics[width=50mm]{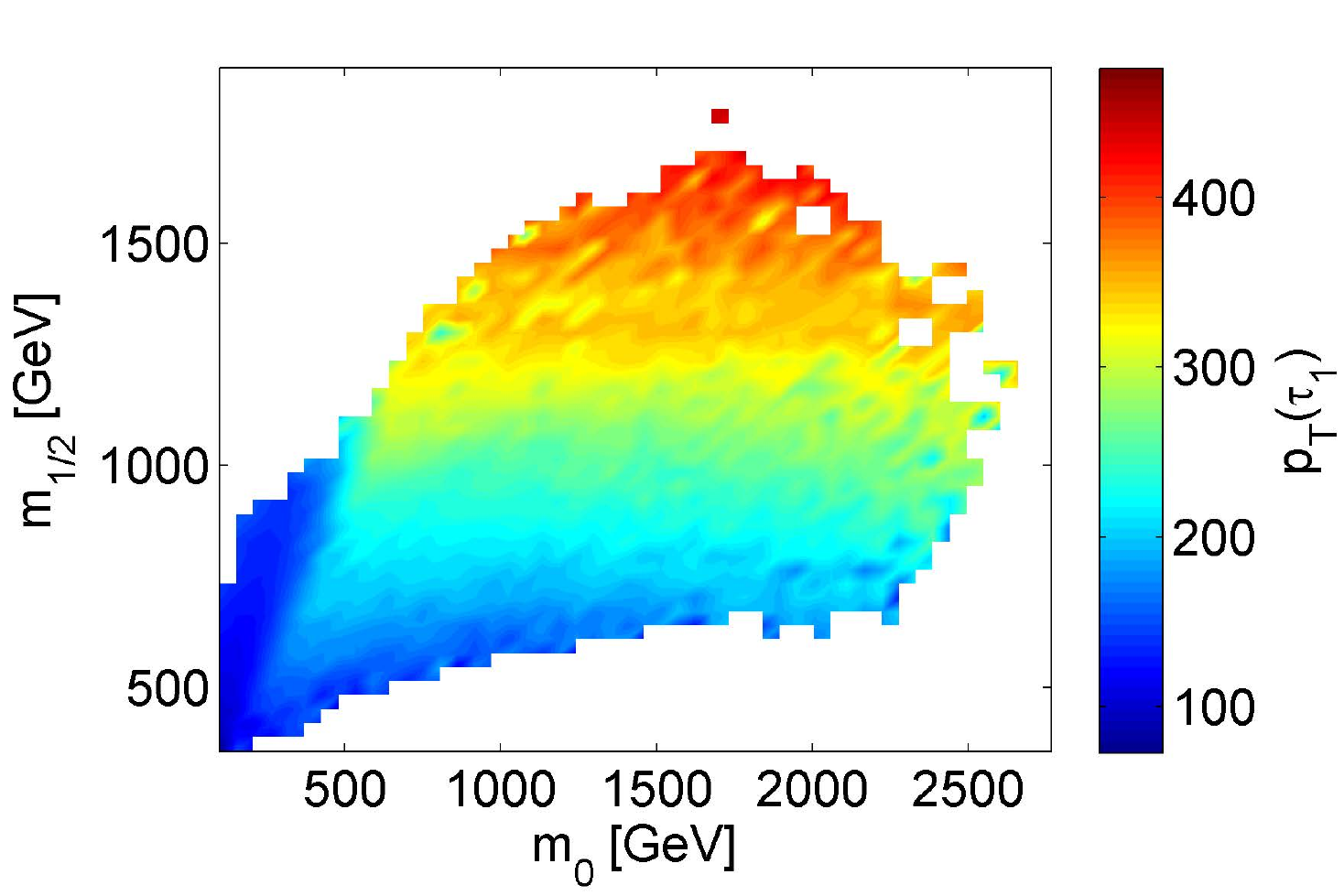}}
   &
   \subfloat{\includegraphics[width=50mm]{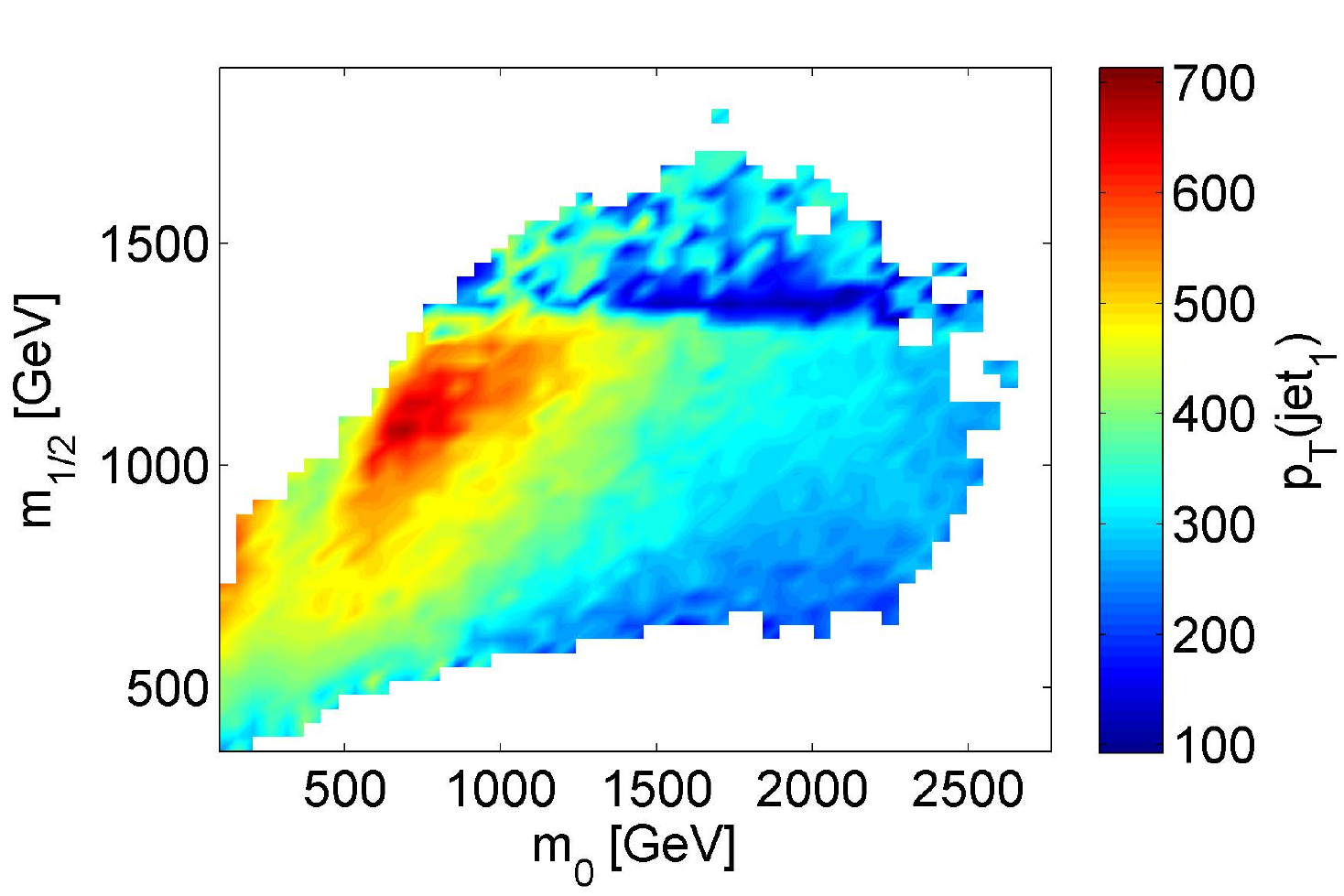}}
   &
   \subfloat{\includegraphics[width=50mm]{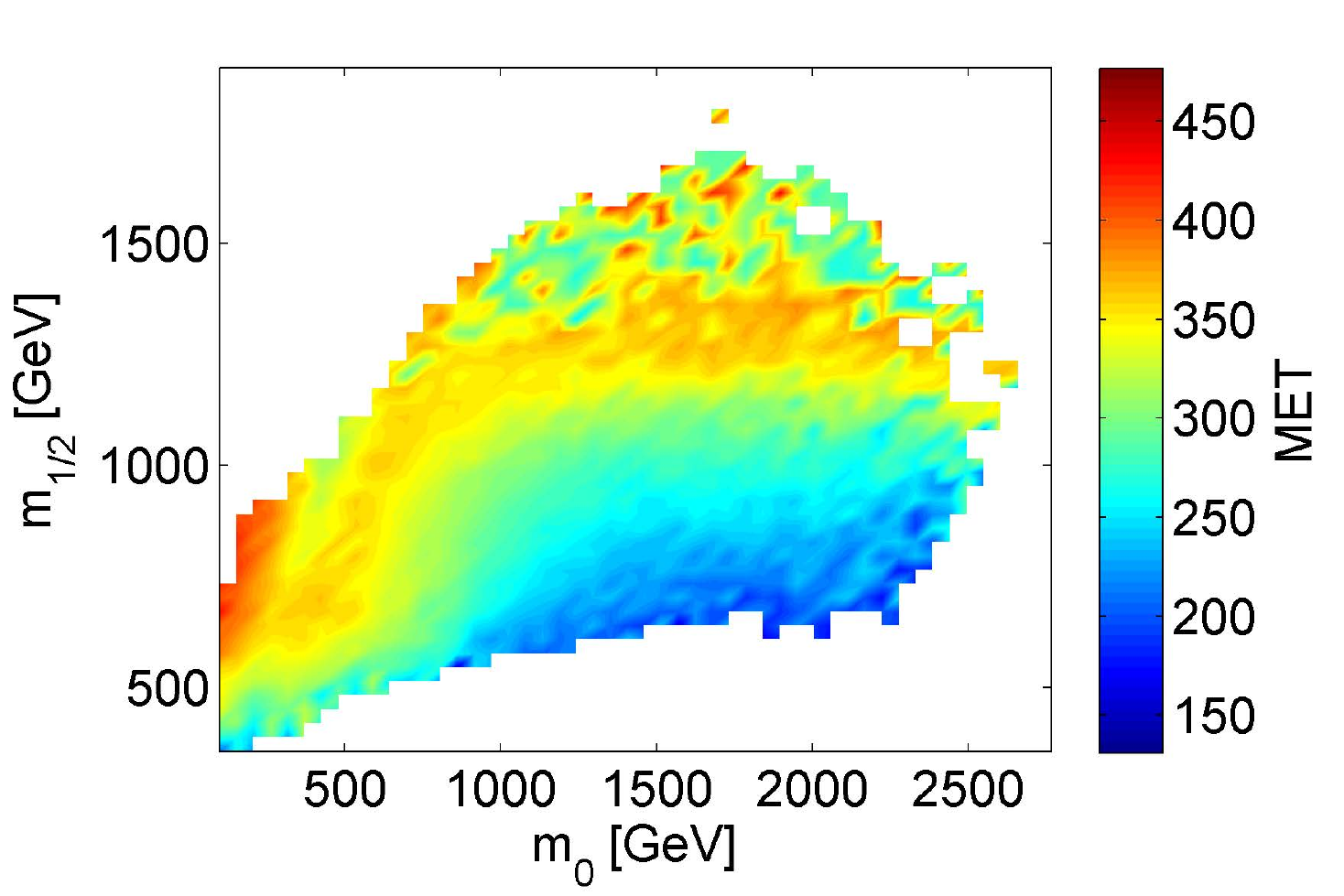}}
   \\
   \subfloat[$p_T(\tau_1)$]{\includegraphics[width=50mm]{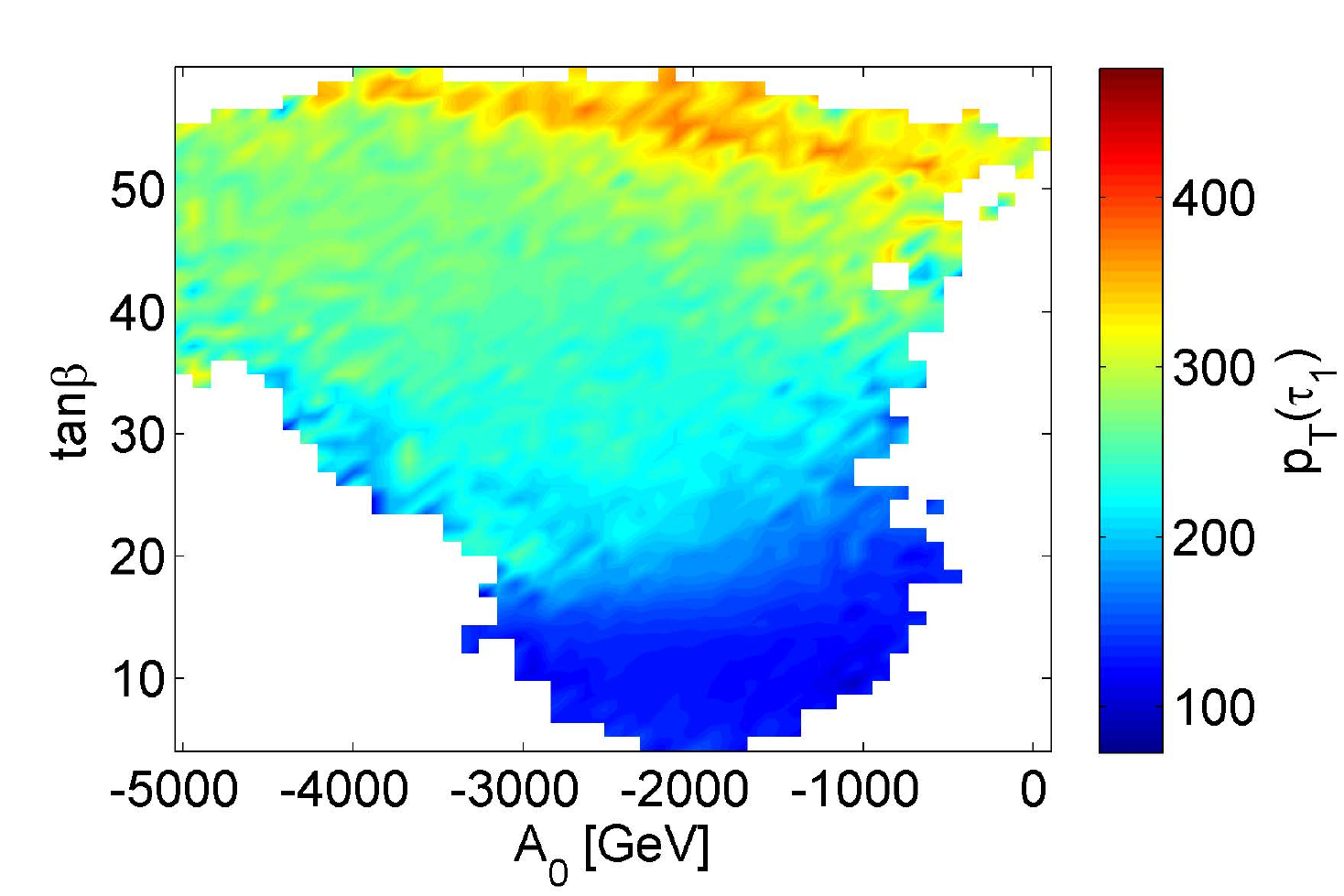}}
   &
   \subfloat[$p_T(jet_1)$]{\includegraphics[width=50mm]{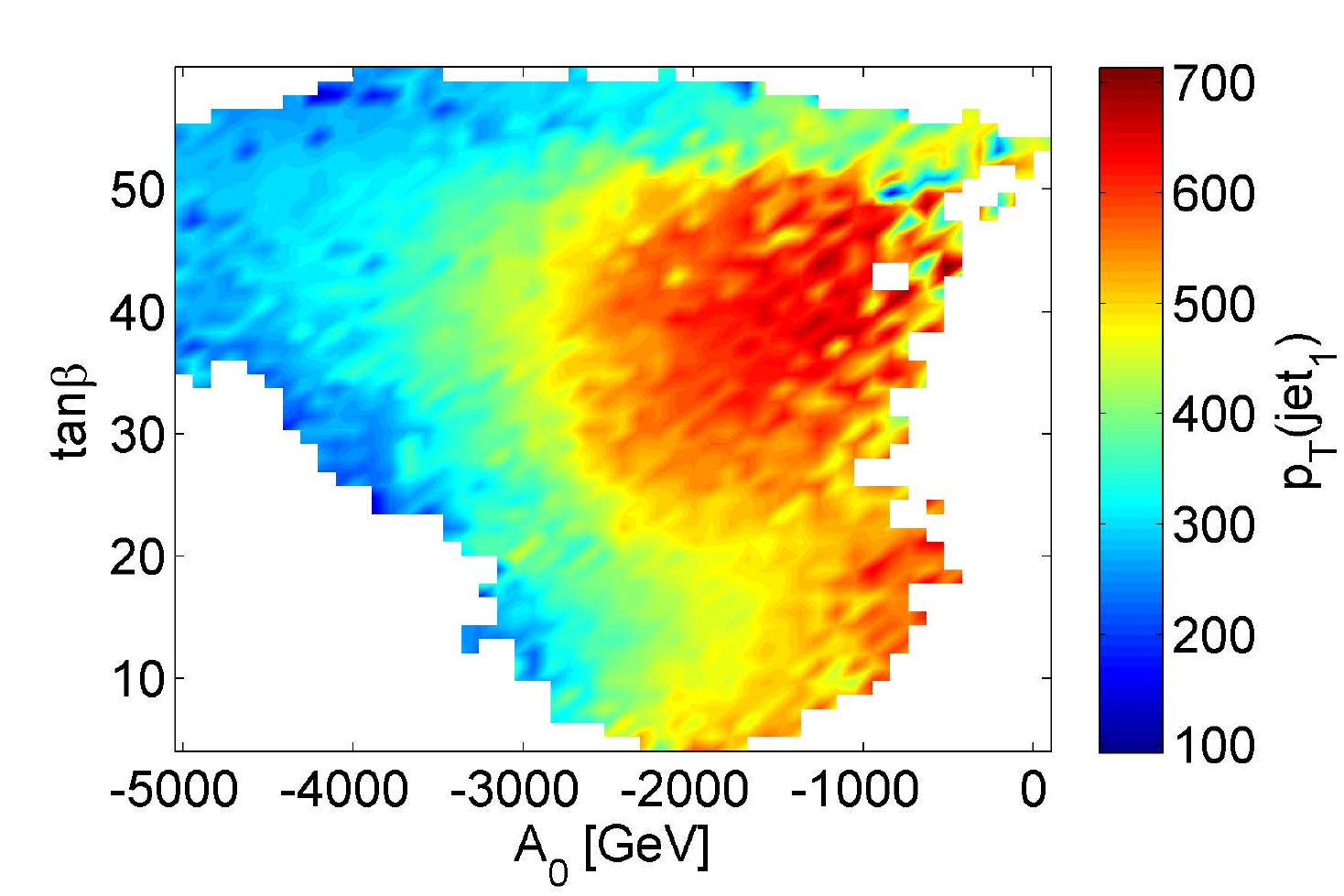}}
   &
   \subfloat[$\slashed E_T$]{\includegraphics[width=50mm]{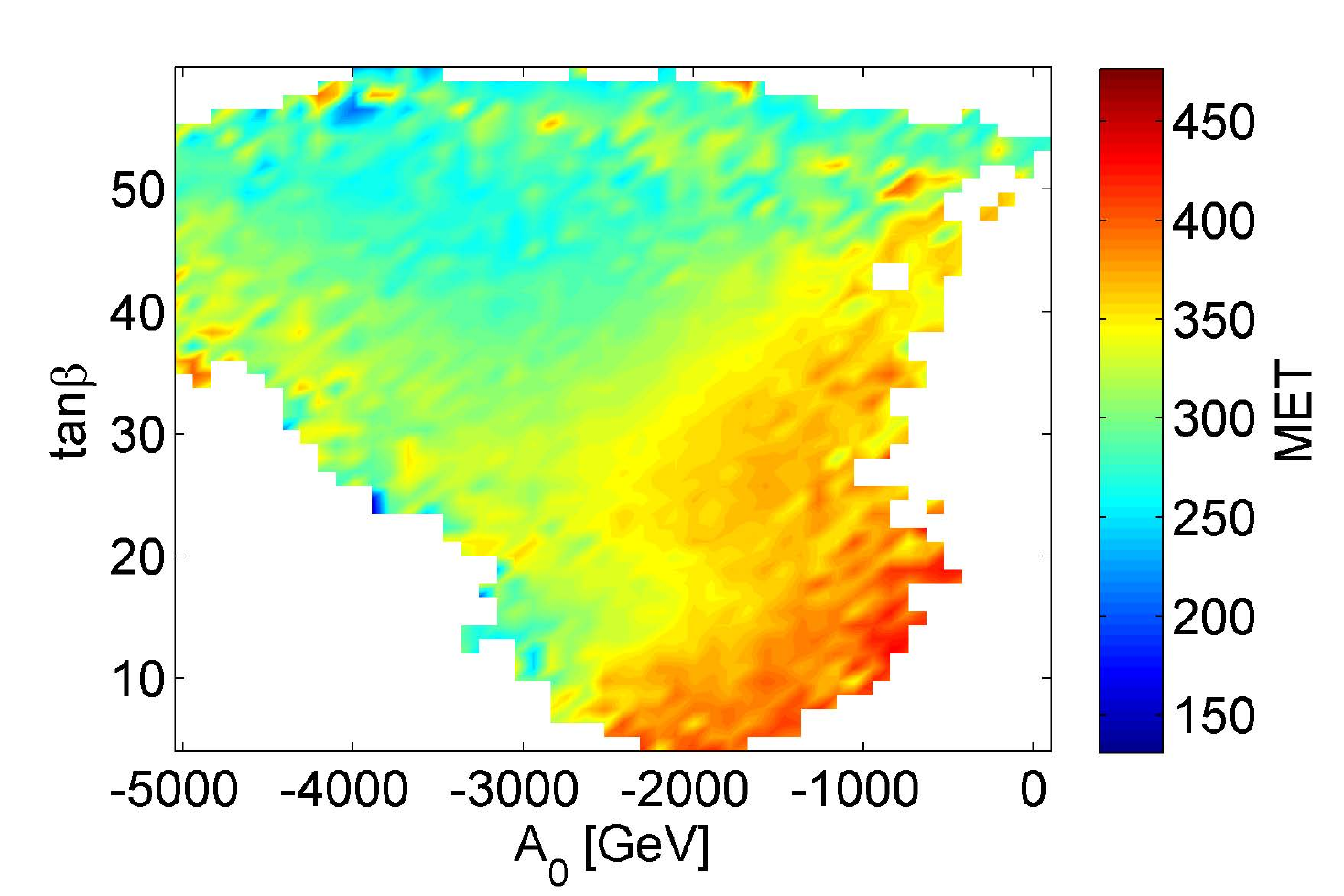}}
   \end{tabular}}
 \end{center}
 \caption{Average value per bin for mean $\slashed E_T$ and $p_T$ for leading jet and $\tau$ shown in the CMSSM mass plane $m_0-m_{1/2}$ (above) and $A_0-\tan\beta$ plane (below)}
 \label{fig:muPhenopT} 
\end{figure}

\begin{figure}[htbp]
 \begin{center}
   \makebox[\textwidth]{
   \begin{tabular}{cc}
   
   \subfloat[$m_0$ vs $m_{1/2}$ plane.]{\includegraphics[width=80mm]{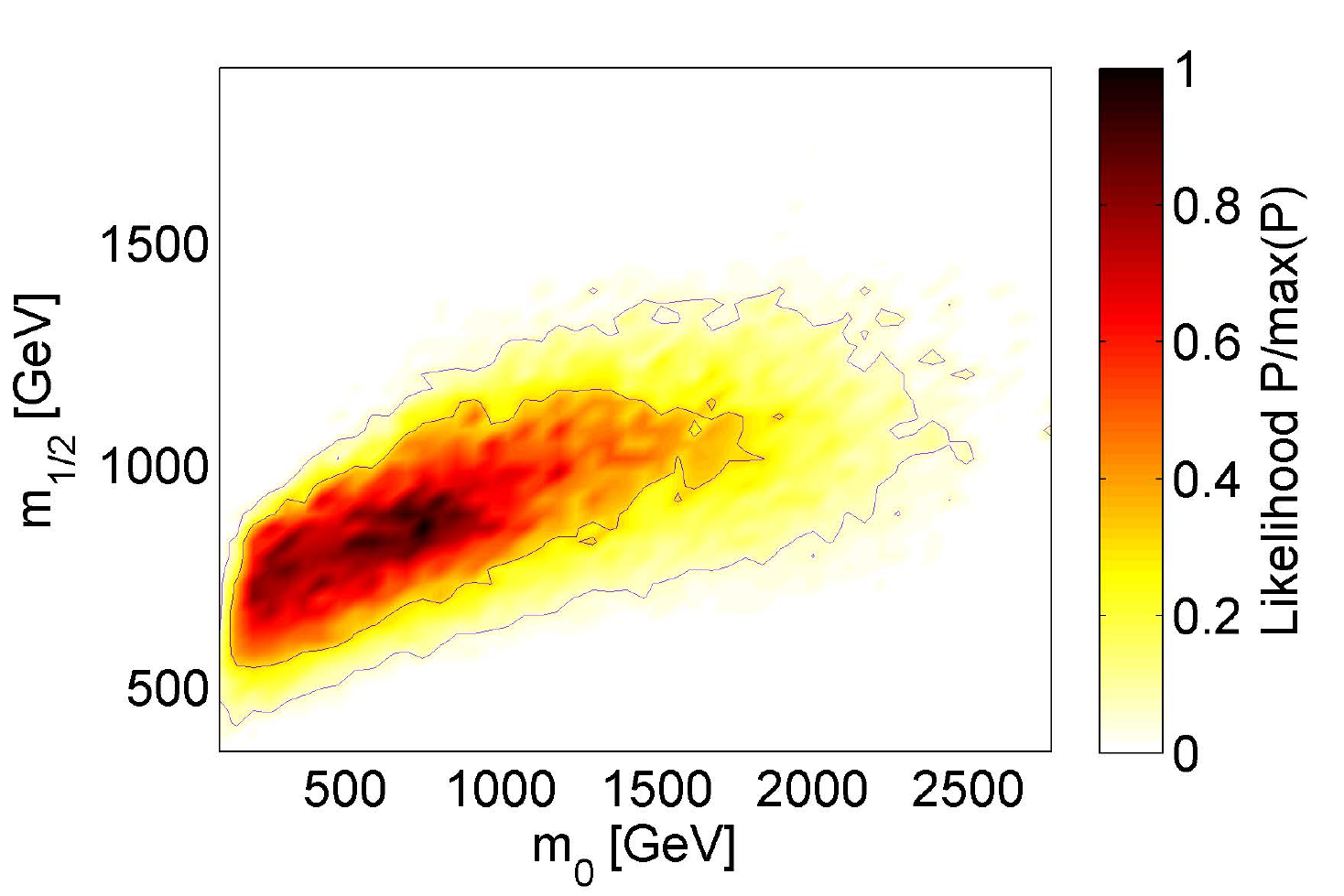}}
   &
   \subfloat[$A_0$ vs $\tan\beta$ plane.]{\includegraphics[width=80mm]{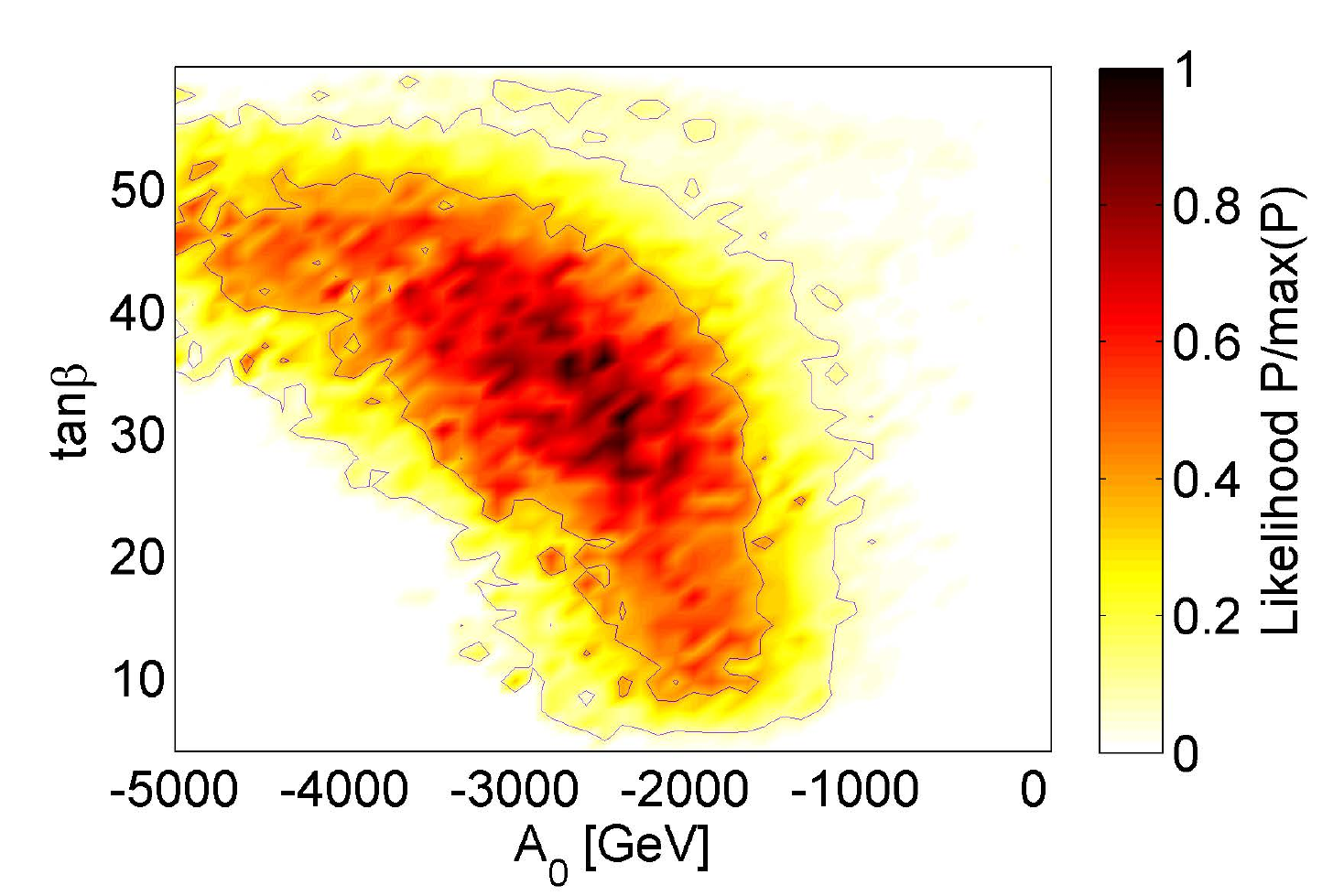}}
   \\
   \subfloat[$m_0$ vs $A_0$ plane.]{\includegraphics[width=80mm]{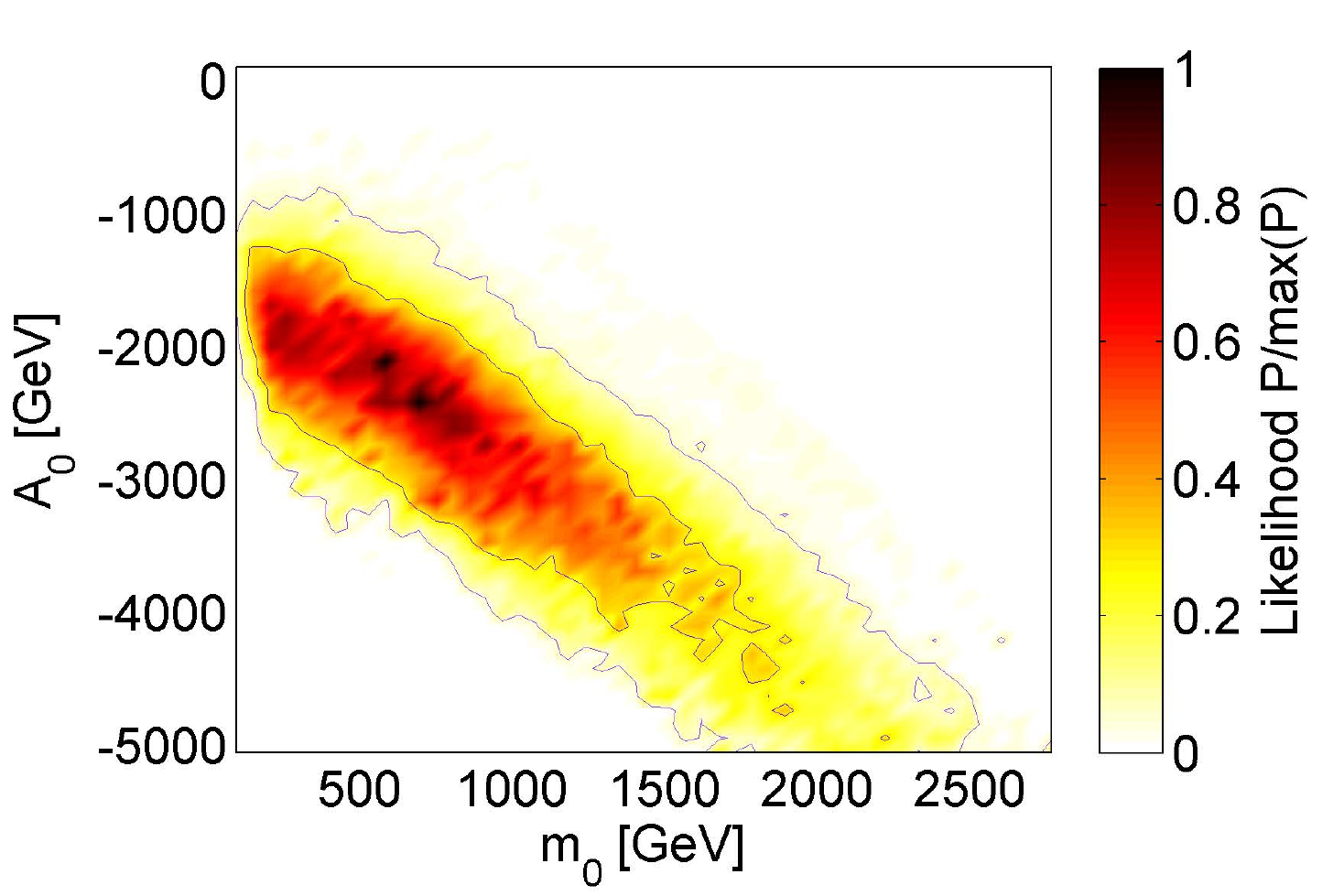}}
   &
   \subfloat[$m_0$ vs $\tan\beta$ plane.]{\includegraphics[width=80mm]{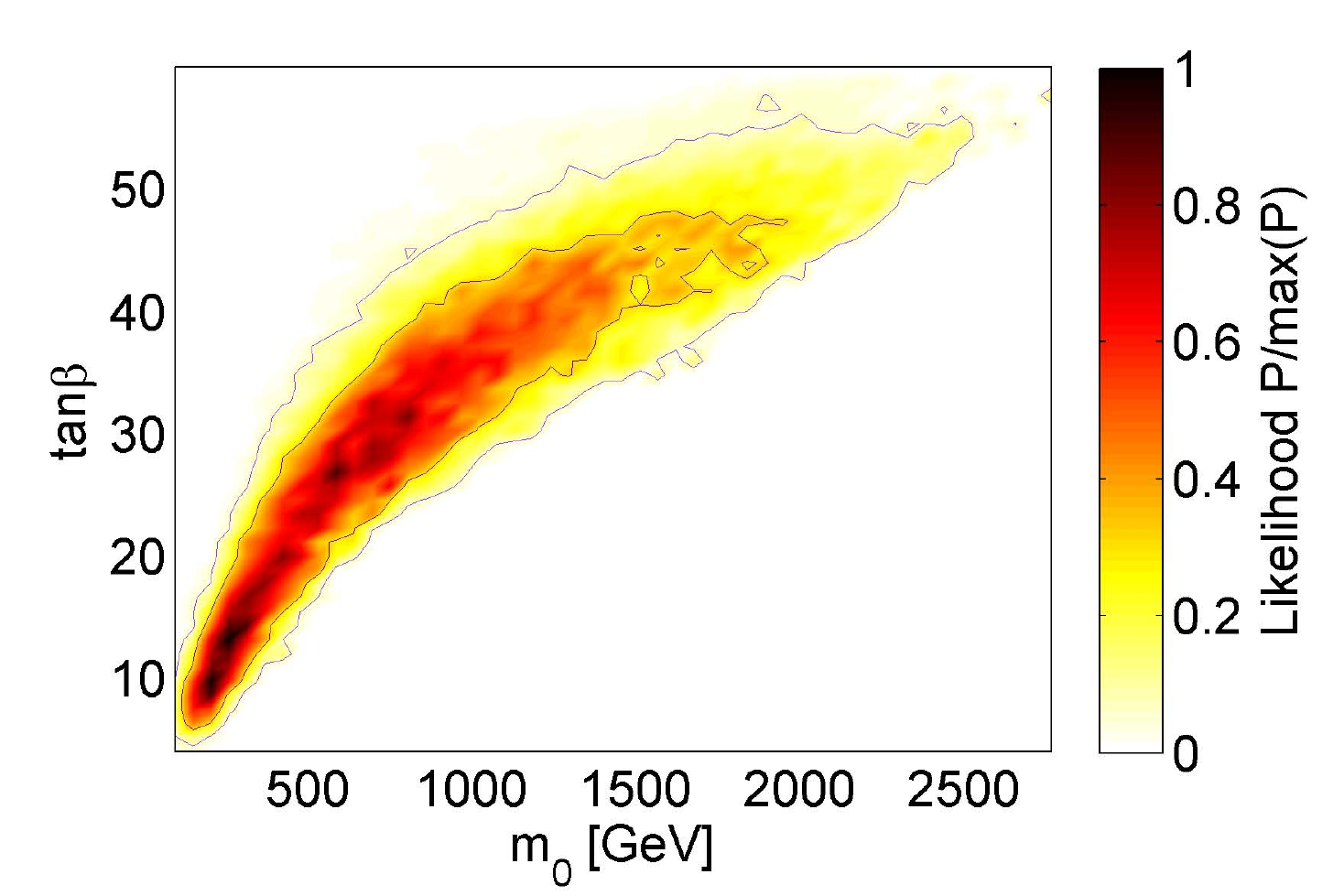}}
   \\
   \subfloat[$m_0$ vs $A_0$ plane.]{\includegraphics[width=80mm]{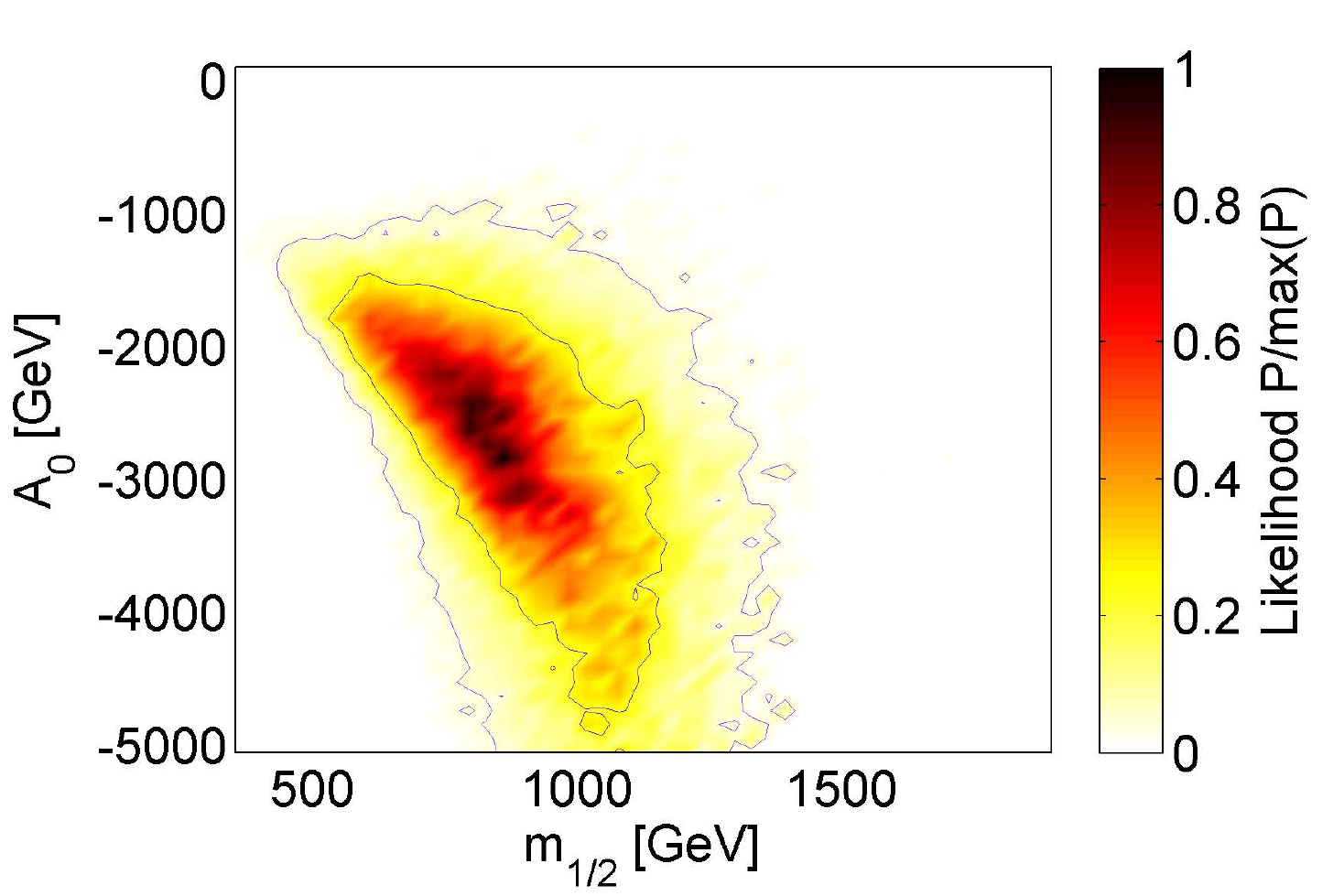}}
   &
   \subfloat[$m_{1/2}$ vs $\tan\beta$ plane.]{\includegraphics[width=80mm]{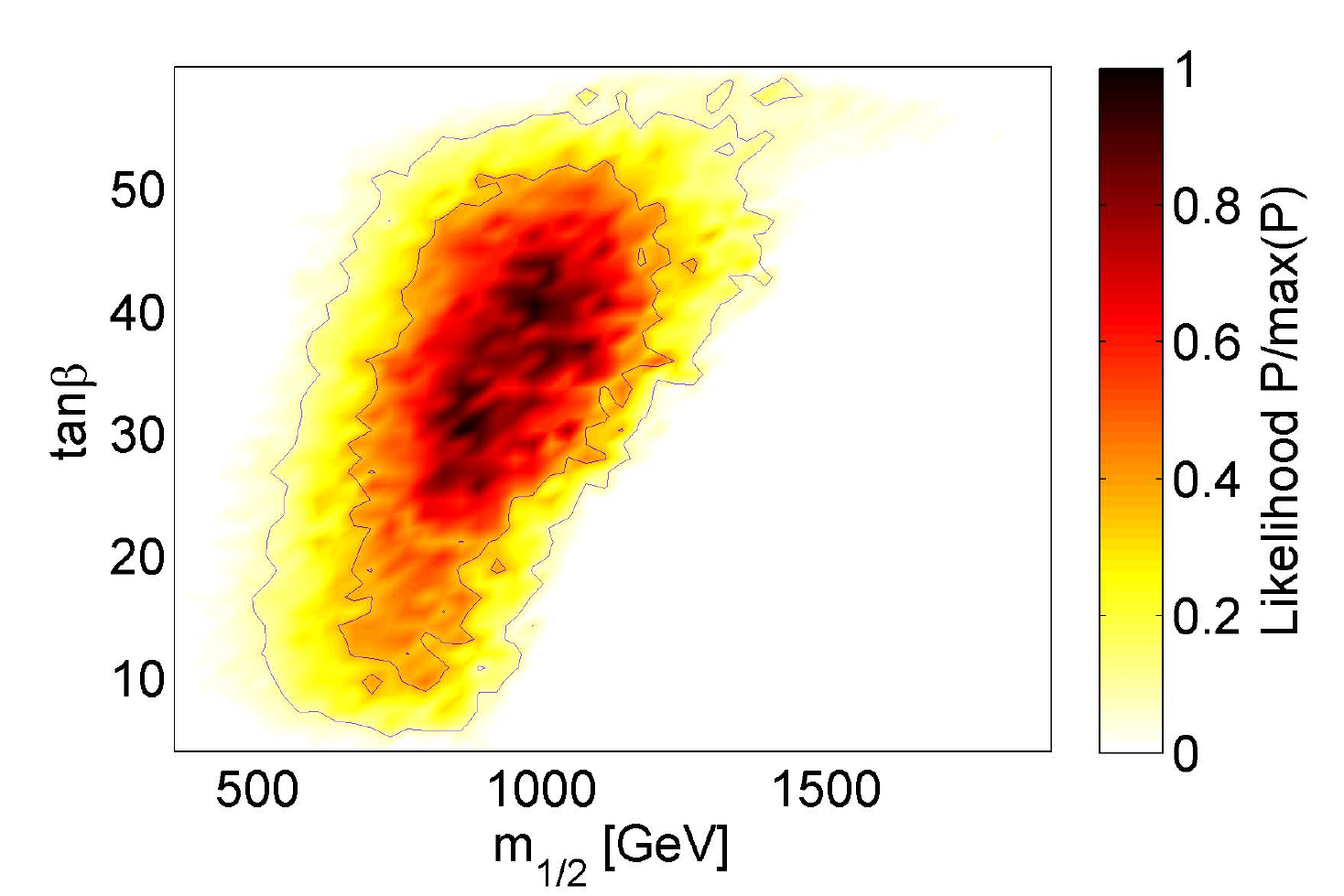}}
   
   \end{tabular}}
 \end{center}
 \caption{Marginalized likelihood maps for different planes in CMSSM
   space.}
 \label{fig:LhoodUni} 
\end{figure}

All viable models found have large negative values of $A_0$ in common, but otherwise span a 
relatively large range of sparticle masses and values of $\tan\beta$. 
The ranges for the  mean $\slashed E_T$ and $p_T$ for leading jet and $\tau$ 
are shown in the CMSSM mass planes $m_0-m_{1/2}$  and $A_0-\tan\beta$ in
figure~\ref{fig:muPhenopT} and  
two dimensional likelihood distributions 
are shown in figure~\ref{fig:LhoodUni}. The distributions are constructed by binning the models 
into $N_\mathrm{bins}=50$ bins along each dimension, where the likelihood of each bin is
 approximated by the number 
of models contained. The effects of different constraints separately are discussed in the 
Appendix~\ref{sec:appendix}. 

\newpage
The {\it discoverability} likelihood
constrains the SUSY $\tau$ production cross-section to not be too small.
This sets upper bounds on how large the gaugino and scalar masses 
can be since the production cross-section falls sharply as the masses of colored sparticles
grow. The cross-section has as well a 
slight $A_0$ dependence which allows for higher masses at higher negative values of $A_0$.

\begin{figure}[htbp]
 \begin{center}
   \makebox[\textwidth]{
   \begin{tabular}{cc}
   
   \subfloat{\includegraphics[width=80mm]{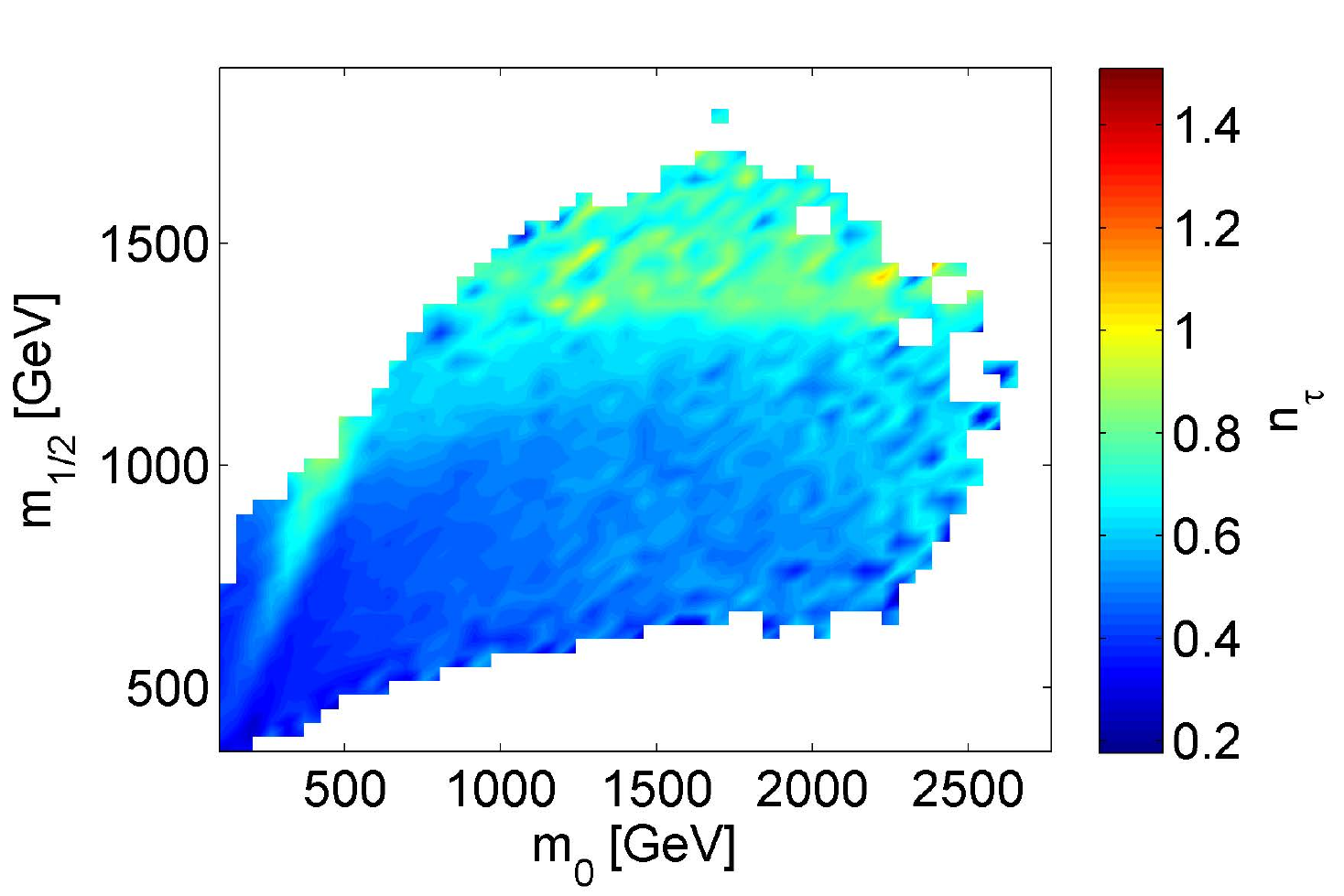}}
   &
   \subfloat{\includegraphics[width=80mm]{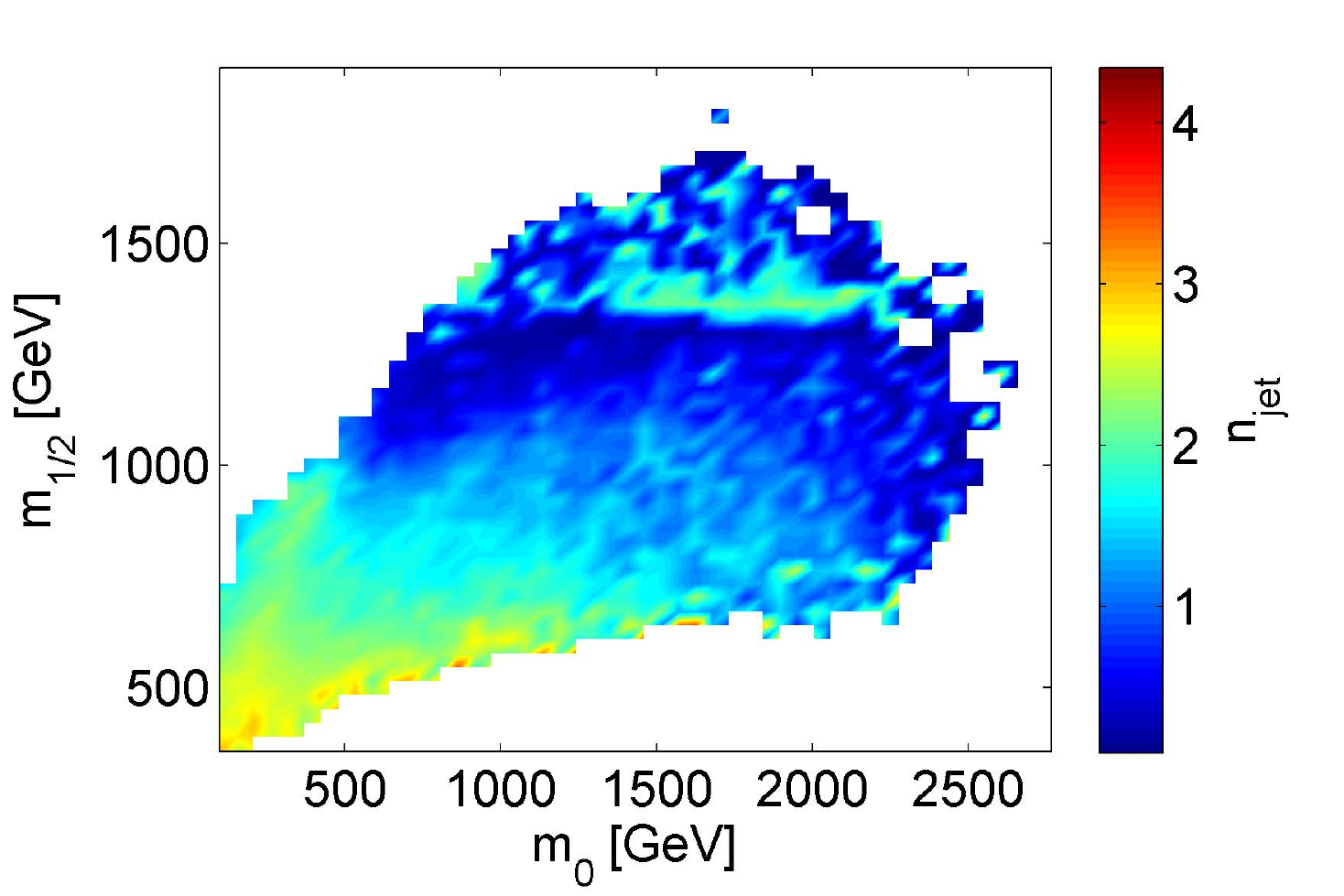}}
   \\
   \subfloat[Mean number of $\tau$s per SUSY event, $n_\tau$]{\includegraphics[width=80mm]{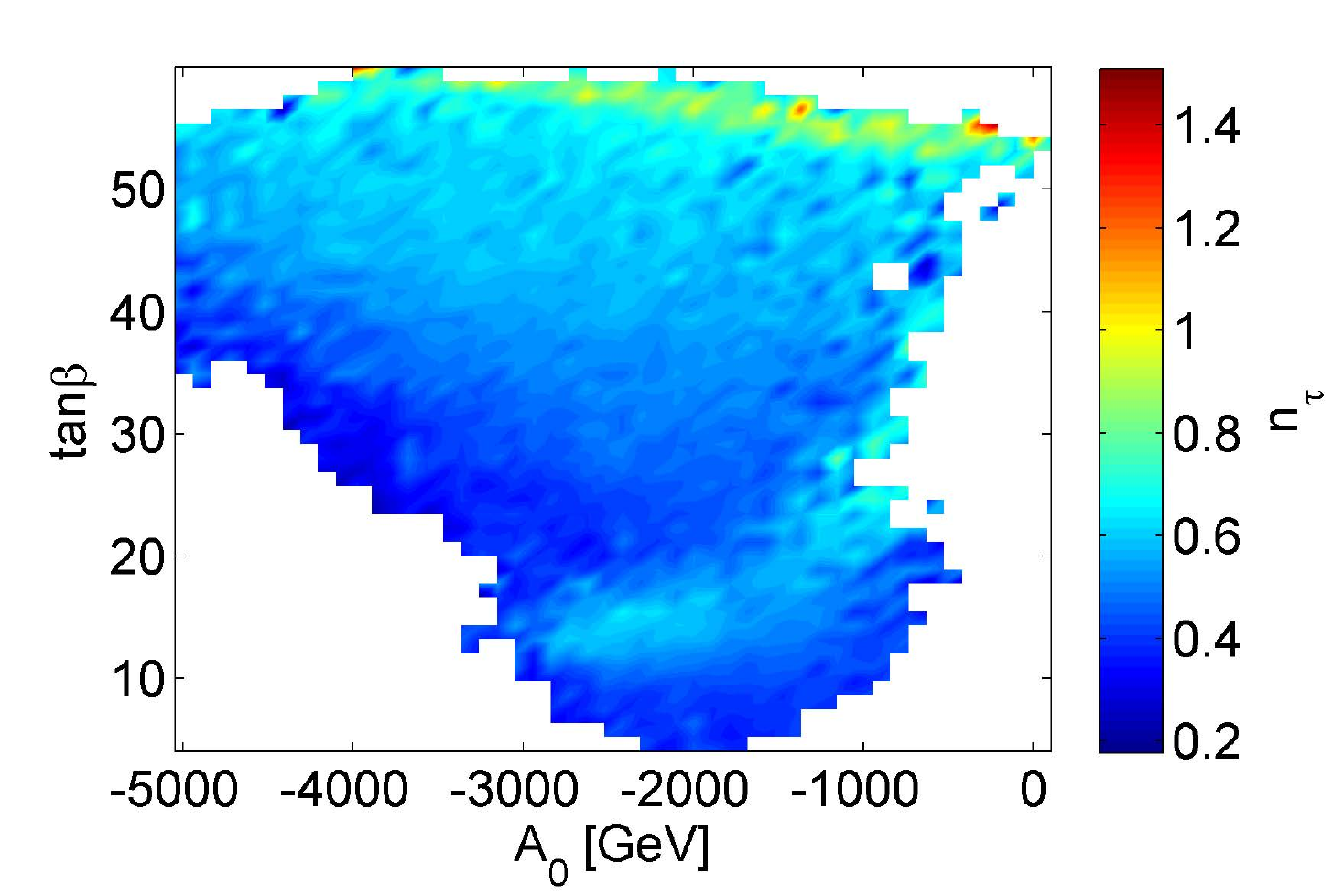}}
   &
   \subfloat[Mean number of jets per SUSY event, $n_{jet}$]{\includegraphics[width=80mm]{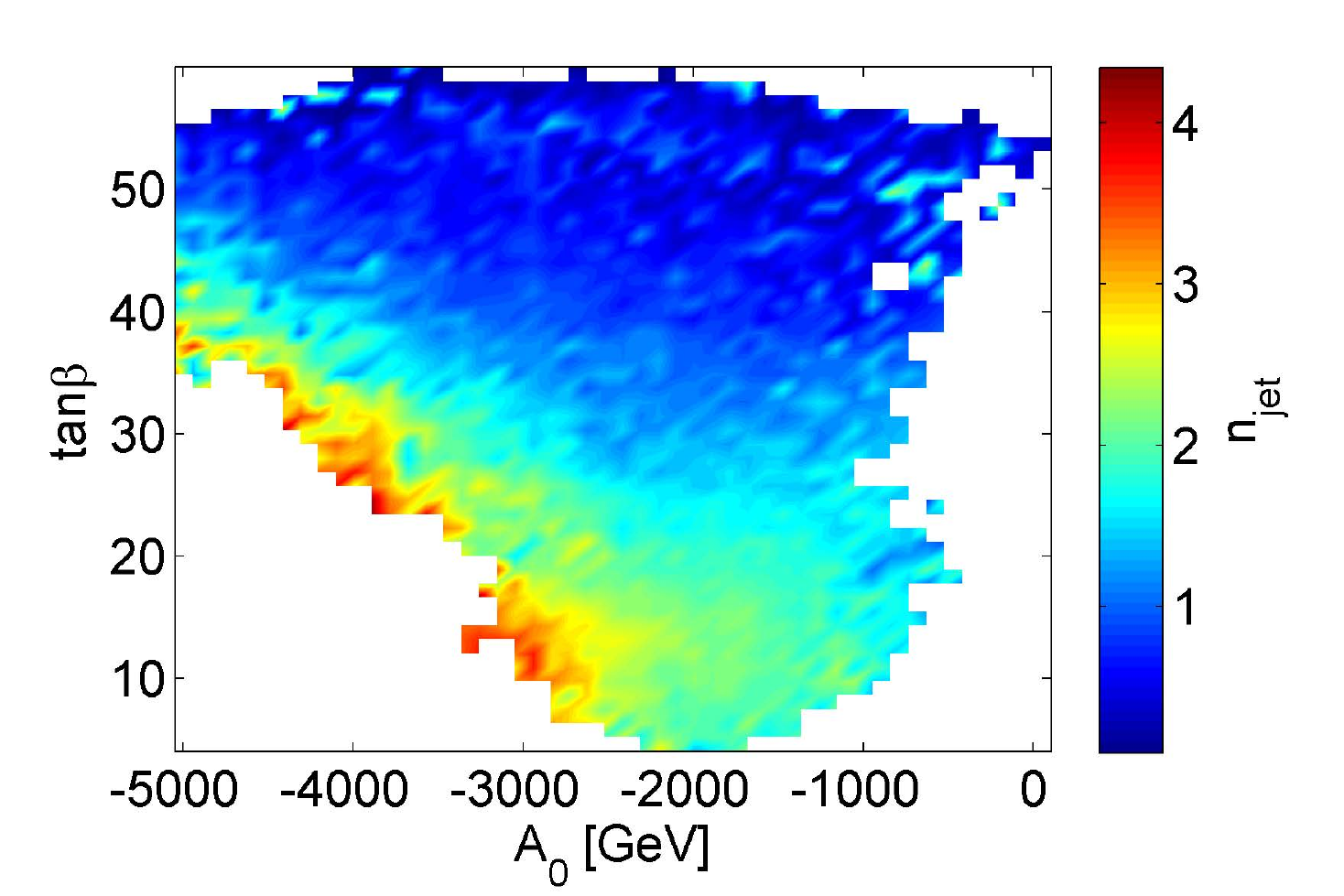}}
   \end{tabular}}
 \end{center}
 \caption{Average value per bin for the mean number of $\tau$s and jets per SUSY event shown in the  $m_0-m_{1/2}$ plane (above) and $A_0-\tan\beta$ plane (below)}
 \label{fig:muPhenon} 
\end{figure}         

\subsection{Phenomenology and Reference Points}
\label{sec:Reference}

The relatively wide range of values for SUSY masses and values of $\tan\beta$ found leads to a
 wide range of values of phenomenological properties such as average 
$\slashed E_T$,the average missing energy per SUSY event, $p_T(\tau_1)$, $p_T(\mathrm{jet}_1)$, the average $p_T$ 
of the leading $\tau$ and the leading jet~ see figure~\ref{fig:muPhenopT}, and $n_{\tau}$, $n_\mathrm{jet}$, the average number of $\tau$'s/jets per 
SUSY event, see figure~\ref{fig:muPhenon}. 

\renewcommand*\arraystretch{1.5}
\begin{table}[htbp]
\caption{The range and best fit value for CMSSM parameters, relic density, $m_{h0}$, expected number of events with taus, $\left<N_{\tau}\right>$ and $\textbf{Br}_{B_s\rightarrow \mu\mu}$.}
\label{tab:BFRange}
  \begin{center}
  \makebox[\textwidth]{
\begin{tabular}{|ccc|ccc|ccc|ccc|}\hline min & \textbf{bf} & max & min& \textbf{bf} & max  & min& \textbf{bf} & max & min & \textbf{bf} & max \\\hline\hline 
\multicolumn{3}{|c|}{$\mathbf{m_0}$ \textbf{[GeV]}} &\multicolumn{3}{c|}{$\mathbf{m_{1/2}}$\textbf{[GeV]}}&\multicolumn{3}{c|}{$\mathbf{A_0}$ \textbf{[GeV]}}&\multicolumn{3}{c|}{$\mathbf{\tan\beta}$}\\\hline
$123.7$&$\mathbf{448.8}$&$2739$&$371.3$&$\mathbf{961.1}$&$1881$&$-4998$&$\mathbf{-2673}$&$52.75$&$4.598$&$\mathbf{15.80}$&$59.44$\\\hline\hline
\multicolumn{3}{|c|}{$\mathbf{\Omega h^2}$} &\multicolumn{3}{c|}{$\mathbf{m_{h0}}$\textbf{[GeV]}}&\multicolumn{3}{c|}{$\mathbf{\left<N_{\tau}\right>}$}&\multicolumn{3}{c|}{$\mathbf{Br_{B_s\rightarrow \mu\mu}\ \left[10^{-9}\right]}$}\\\hline
$0.01$&$\mathbf{0.1164}$&$0.1296$&$119.2$&$\mathbf{125}$&$126.2$&$0.01$&$\mathbf{10.64}$&$6784$&$3.840$&$\mathbf{3.907}$&$8.494$\\\hline\hline
\end{tabular}} 
\end{center}
\end{table} 


The $p_T$ values for the leading jet and $\tau$ lepton obviously tend to be higher for high sparticle masses, 
since  higher masses in CMSSM lead to higher mass splittings between the sfermions and the LSP. 
The $p_T$s also become larger with $A_0$ closer to $0$ and for high $\tan\beta$ values. 
The missing energy on the other hand tends to become larger at smaller scalar masses and increasing gaugino mass.
The increase in $\slashed{E_T}$ with $m_{1/2}$ is likely due to increasing neutralino mass. These dependencies are illustrated in figure~\ref{fig:muPhenopT}.

\begin{figure}[htbp]
 \begin{center}
   \makebox[\textwidth]{
   \begin{tabular}{ccc}
   
   \subfloat{\includegraphics[width=80mm]{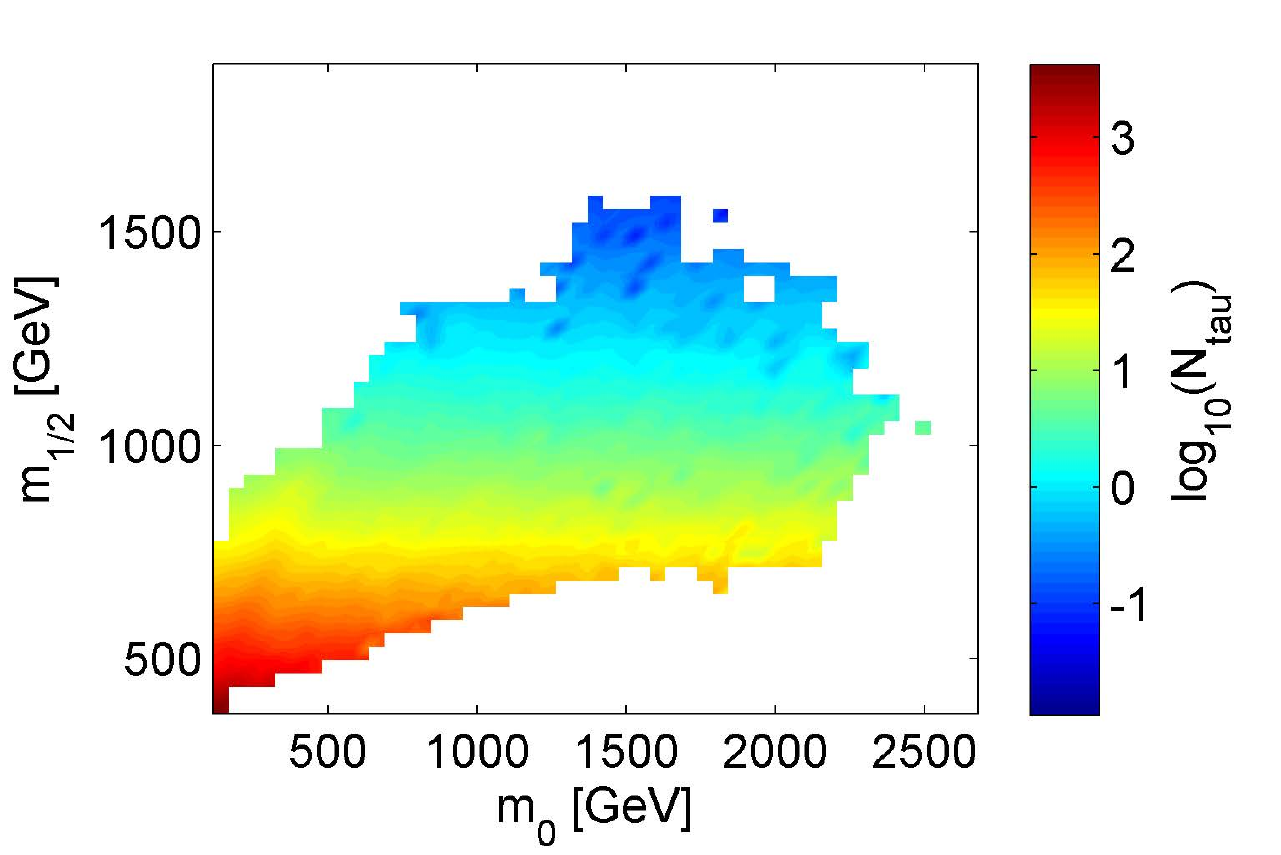}}
   &
   \subfloat{\includegraphics[width=80mm]{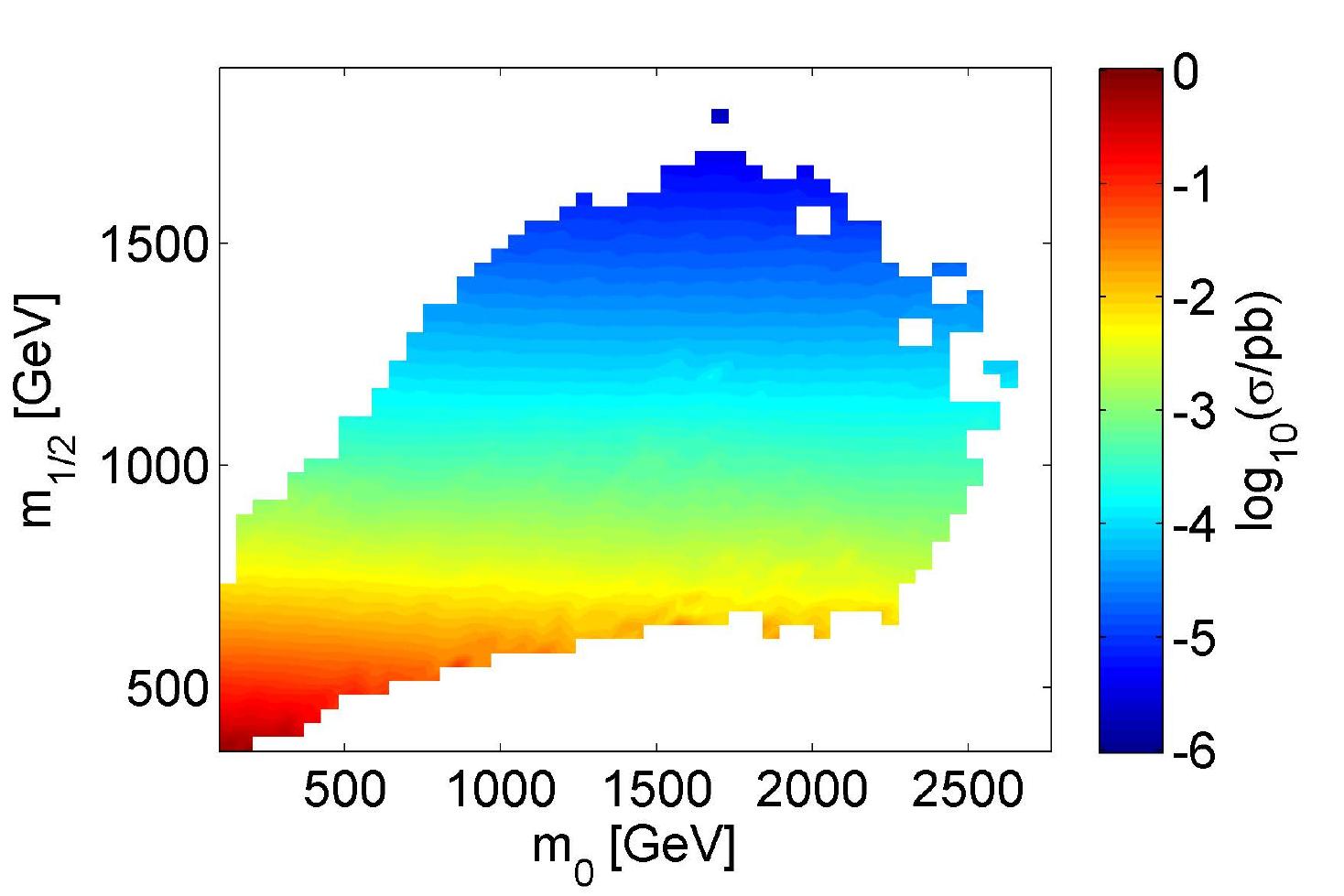}}
   \\
   \subfloat[Expected number of $\tau$ events with $\int\mathcal{L}dt=21\ fb^{-1}$ ]{\includegraphics[width=80mm]{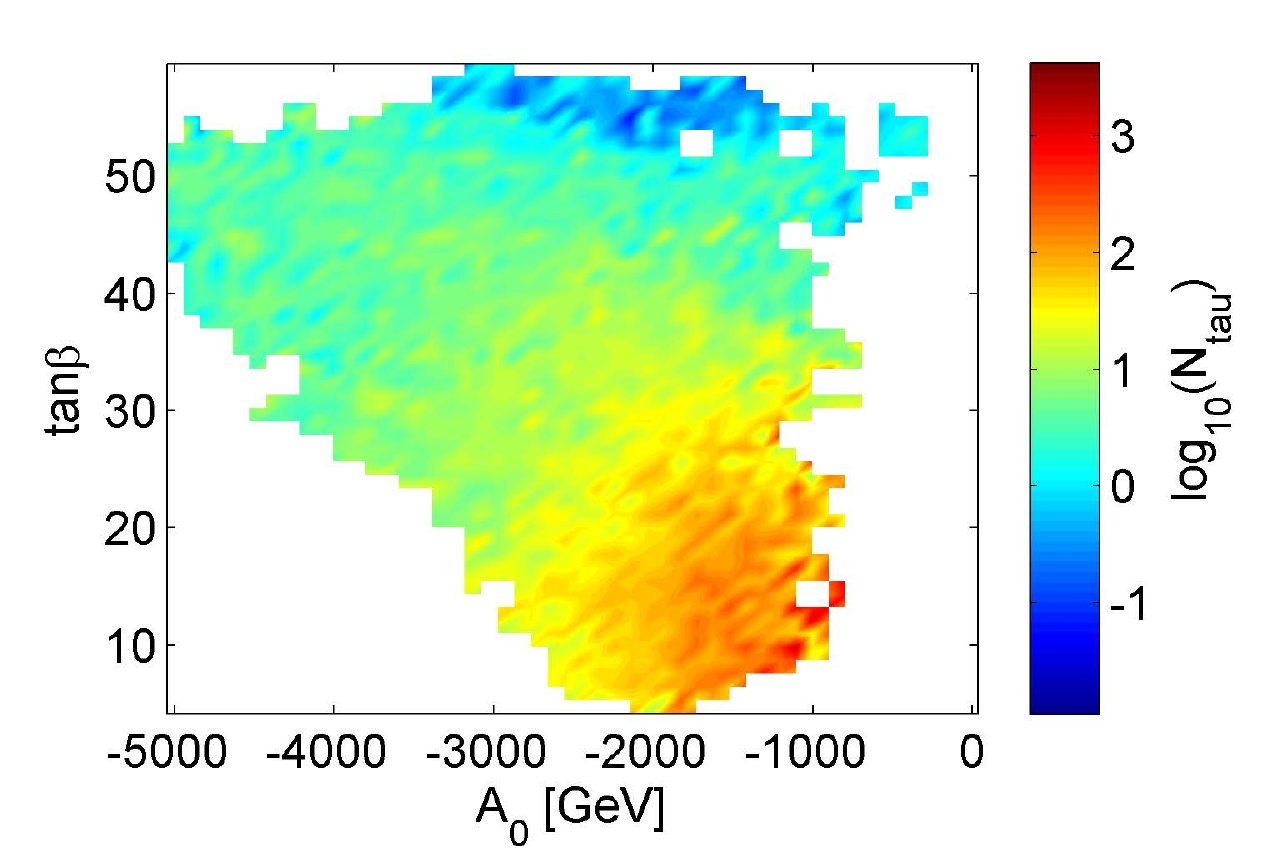}}
   &
   \subfloat[SUSY cross section]{\includegraphics[width=80mm]{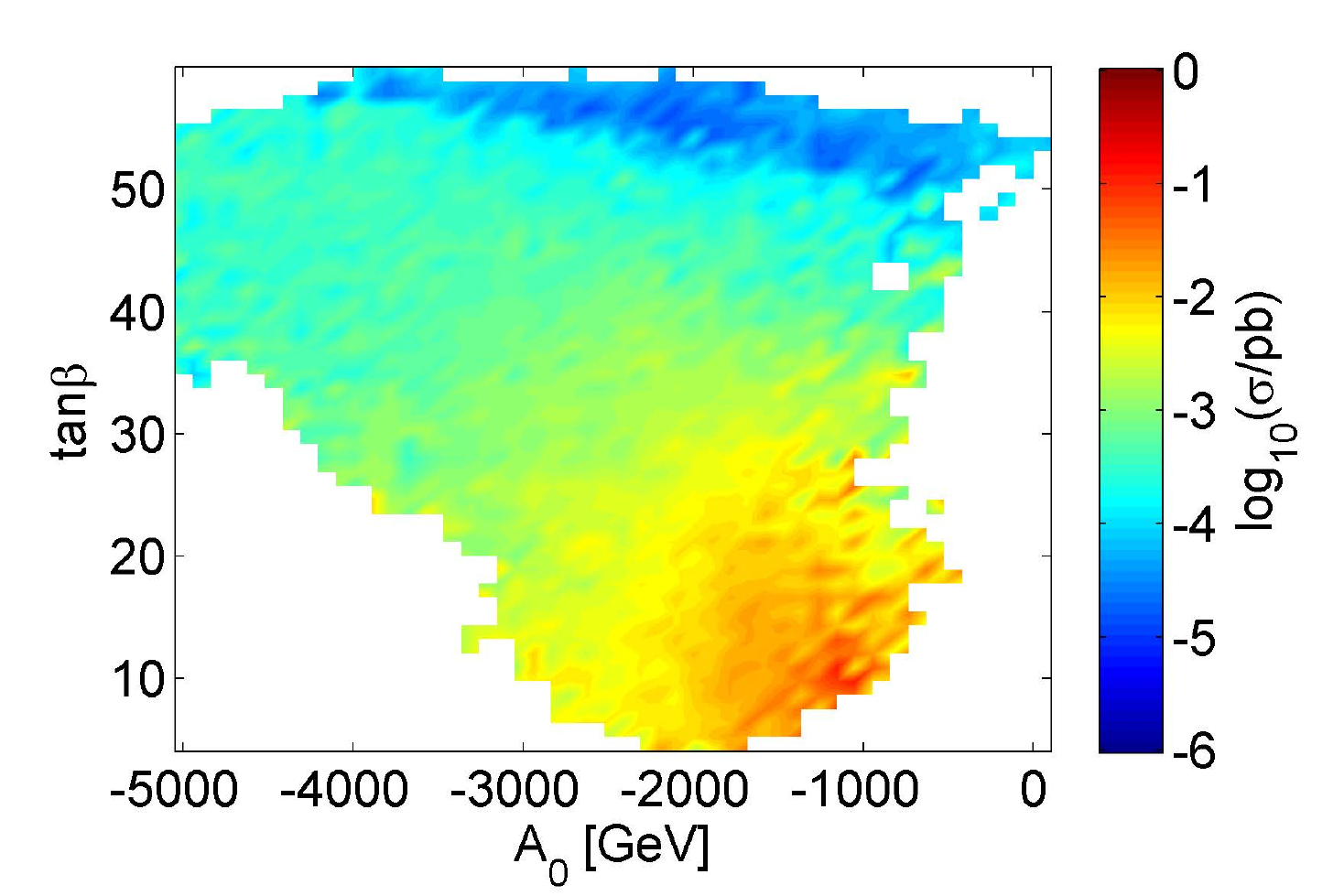}}
   \end{tabular}}
 \end{center}
 \caption{Average value per bin for expected number of $\tau$ events  and SUSY cross-section
 in the mass plane $m_0-m_{1/2}$ (above) and $A_0-\tan\beta$ (below)}
 \label{fig:Ntau} 
\end{figure}

The average number of $\tau$s per SUSY events is mostly due to the branching fraction into $\tau$s as 
it can be seen  comparing figure~\ref{fig:muPhenon} and~\ref{fig:Ntau}.
One tau with high $p_T$  per event is produced on average. 
The SUSY branching fraction to $\tau$s is largest at low values of $\tan\beta$ and $m_0$.
At least one high $p_t$ jet is expected  in almost every event. 
The average numbers of jets increases with $m_0$, $\tan\beta$ and $|A_0|$.  

In order to construct reference benchmark models that cover these different phenomenological 
properties the sample was clustered according to the phenomenological observables: 
$\slashed E_T$, $n_\mathrm{jet}$, $p_T(\mathrm{jet}_1)$, $n_\tau$, $p_T(\tau_1)$, details are described in the Appendix~\ref{sec:appendix}.

With these, ten  phenomenological clusters shown in
table~\ref{tab:clus_pheno} were found.  The SUSY and model related
parameters for these clusters are shown in table~\ref{tab:clus_susy} and ~\ref{tab:clus_spart}.
The centroids of the clusters can be regarded as reference (benchmark) points. additional k-factors were calculated for these benchmark points to give a more precise estimate for the expected number of $\tau$'s. As it can be seen in table~\ref{tab:clus_susy}, NLO corrections are small ranging from $1.1$ to $1.7$, and around 125 events with high energetic taus in the central part of the detector are expected for the highest cross-section reference points, which might be enough to detect the signal using the $21/fb$ of data gathered in 2012. Exploration of the other reference points might have to wait until the LHC 13~TeV operations. 

\begin{figure}[htbp]
  \begin{center}
  \makebox[\textwidth]{
  \begin{tabular}{cc}
  	\subfloat[$m_0$ vs $m_{1/2}$ projection.]{
          \label{fig:sub_susy_clusters_a}
          \includegraphics[width=70mm]{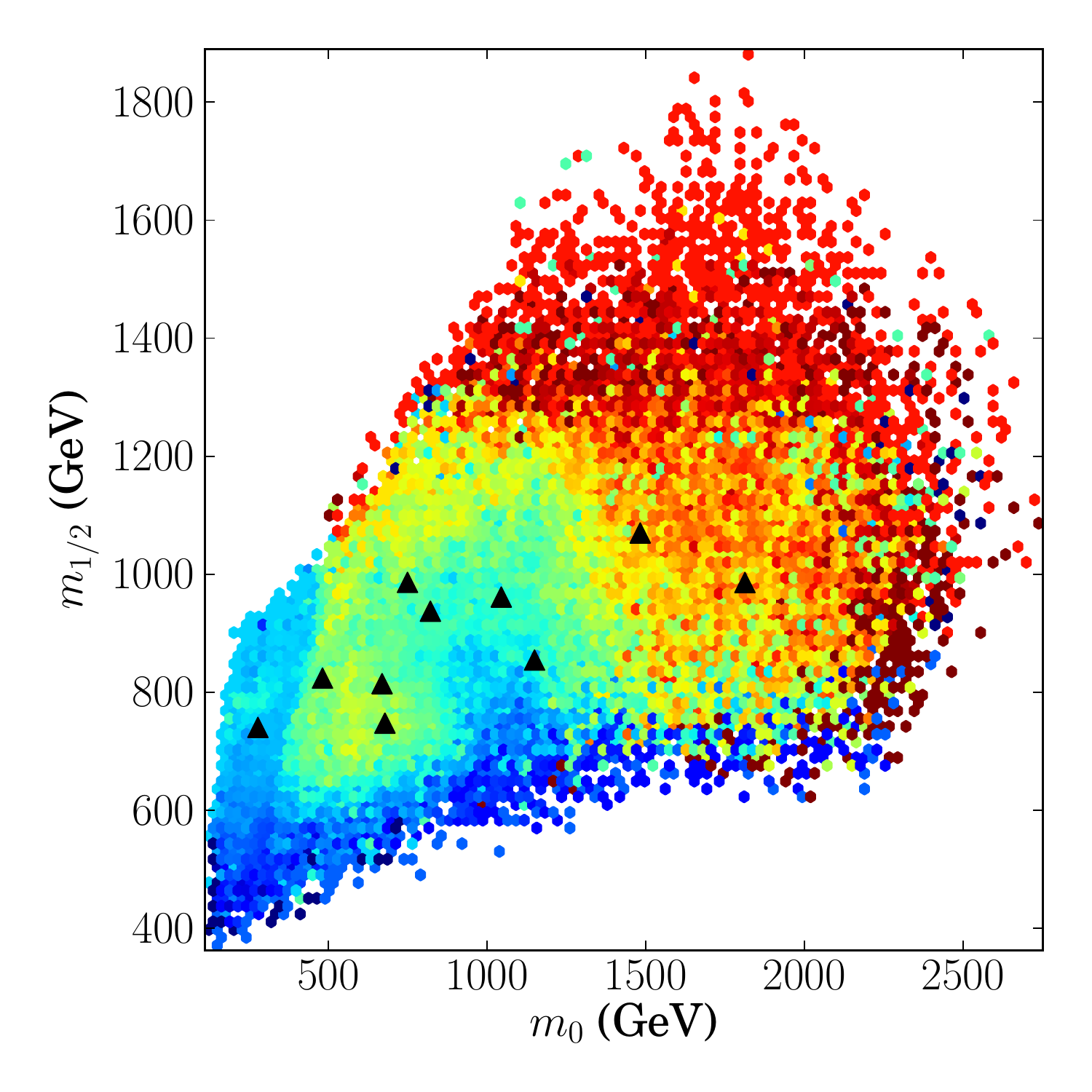}}
  	&    
    \subfloat[$A_0$ vs $\tan\beta$ projection.]{
      \label{fig:sub_susy_clusters_b}
      \includegraphics[width=70mm]{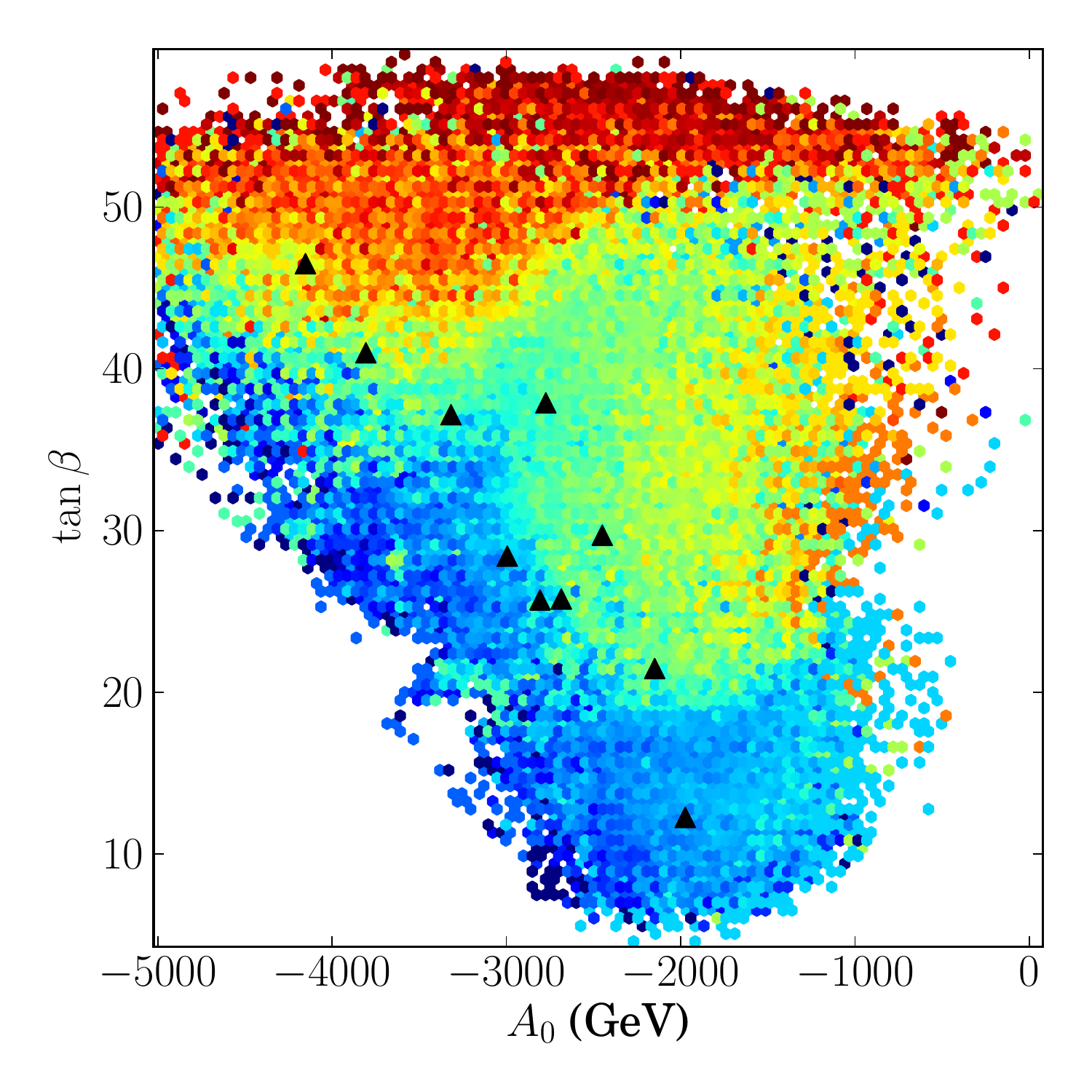}}  
  \end{tabular}}  
  \end{center}
  \caption{Projections of clusters. Colors indicate to
    which cluster a given model belongs.}
  \label{fig:susy_clusters}
\end{figure}

\begin{figure}[htbp]
  \begin{center}
  \makebox[\textwidth]{
  	\begin{tabular}{cc}
  	\subfloat[Leading $\tau$ vs jet $p_T$]{\label{fig:sub_pheno_clusters_a}\includegraphics[width=80mm]{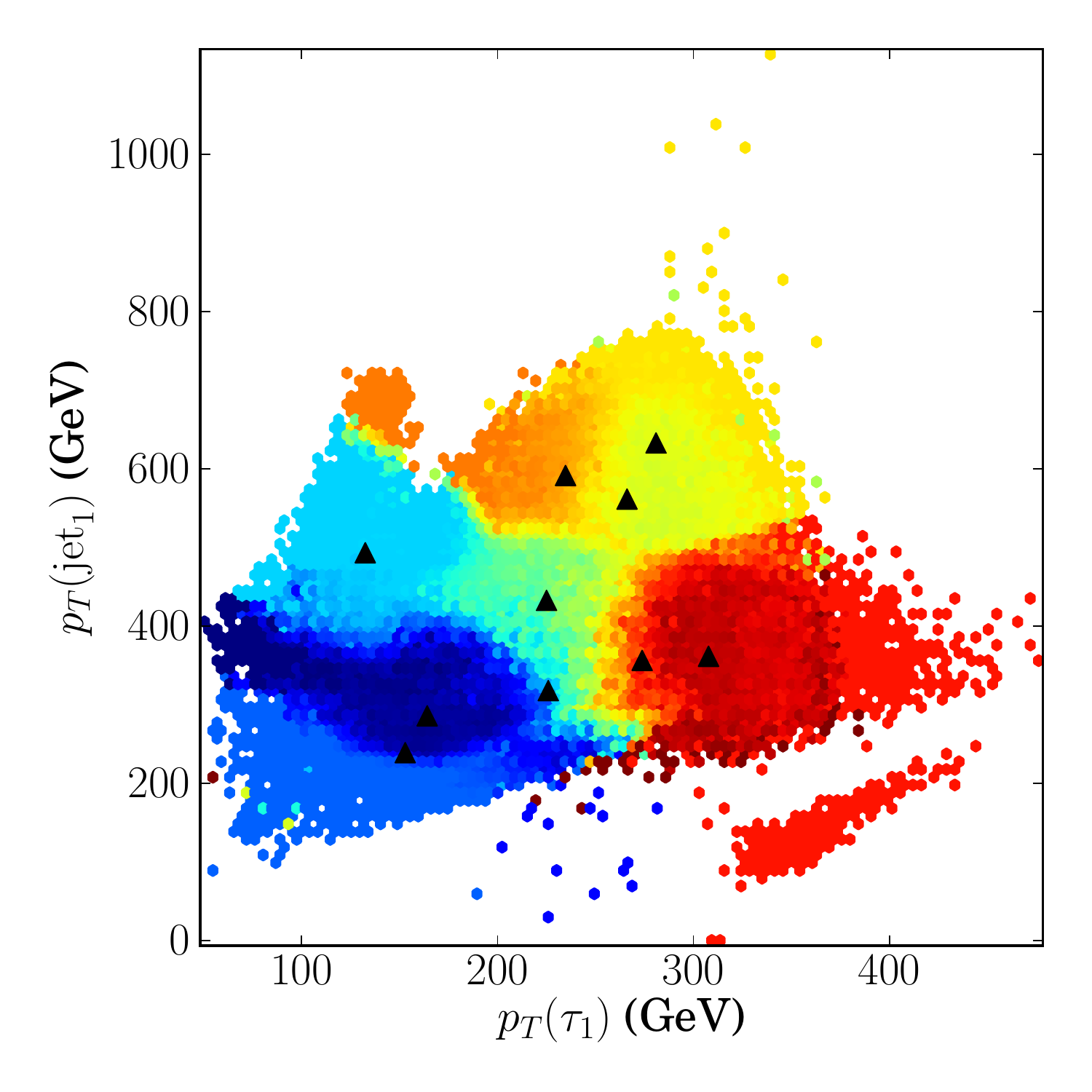}}
  	&
  	\subfloat[$n_\mathrm{jet}$ vs $n_\tau$ projection.]{\label{fig:sub_pheno_clusters_b}\includegraphics[width=80mm]{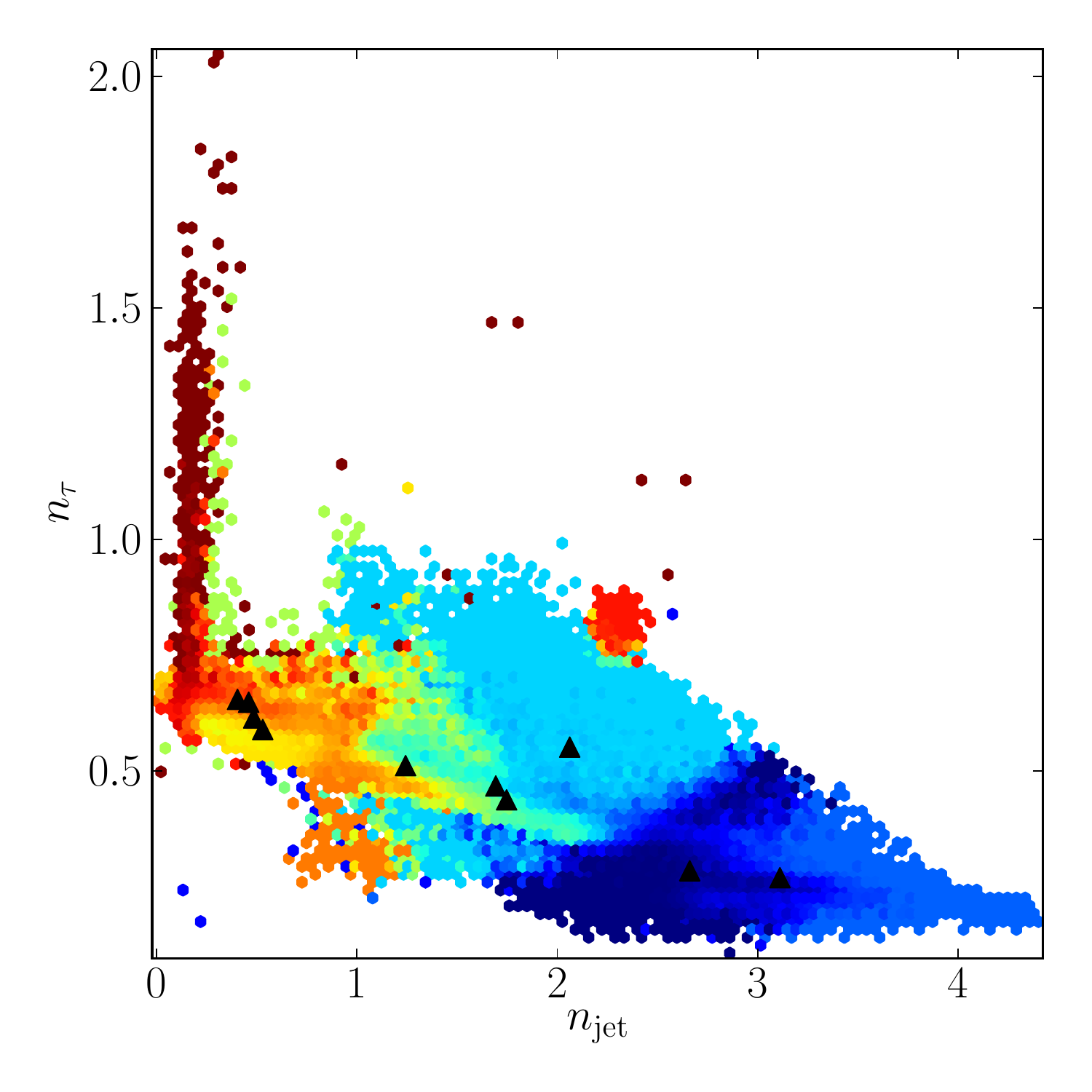}}
  	\\
  	\subfloat[$\slashed E_T$ vs $n_\tau$ projection.]{\label{fig:sub_pheno_clusters_c}\includegraphics[width=80mm]{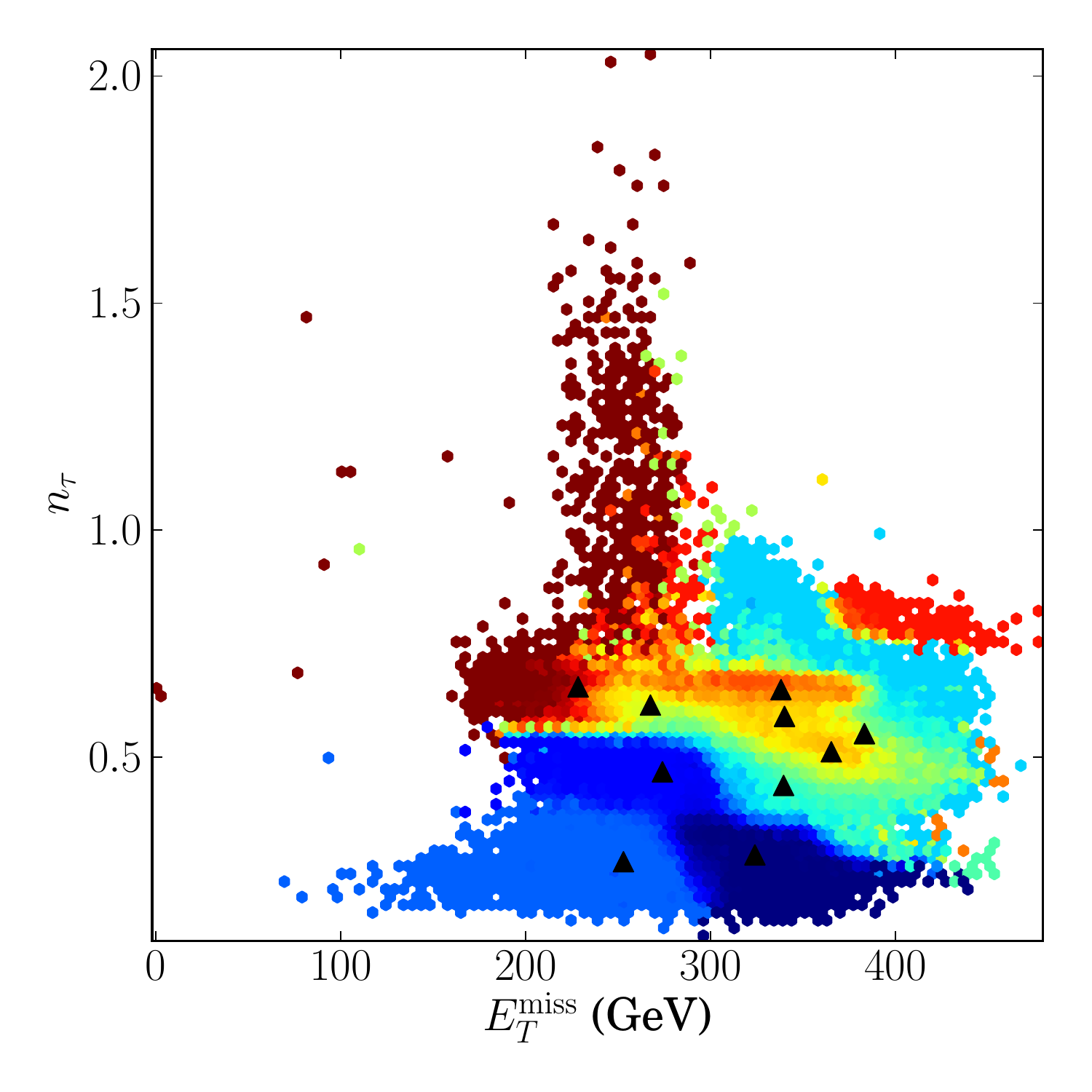}}
    &
    \subfloat[$\slashed E_T$ vs leading jet $p_T$ projection.]{\label{fig:sub_pheno_clusters_e}\includegraphics[width=80mm]{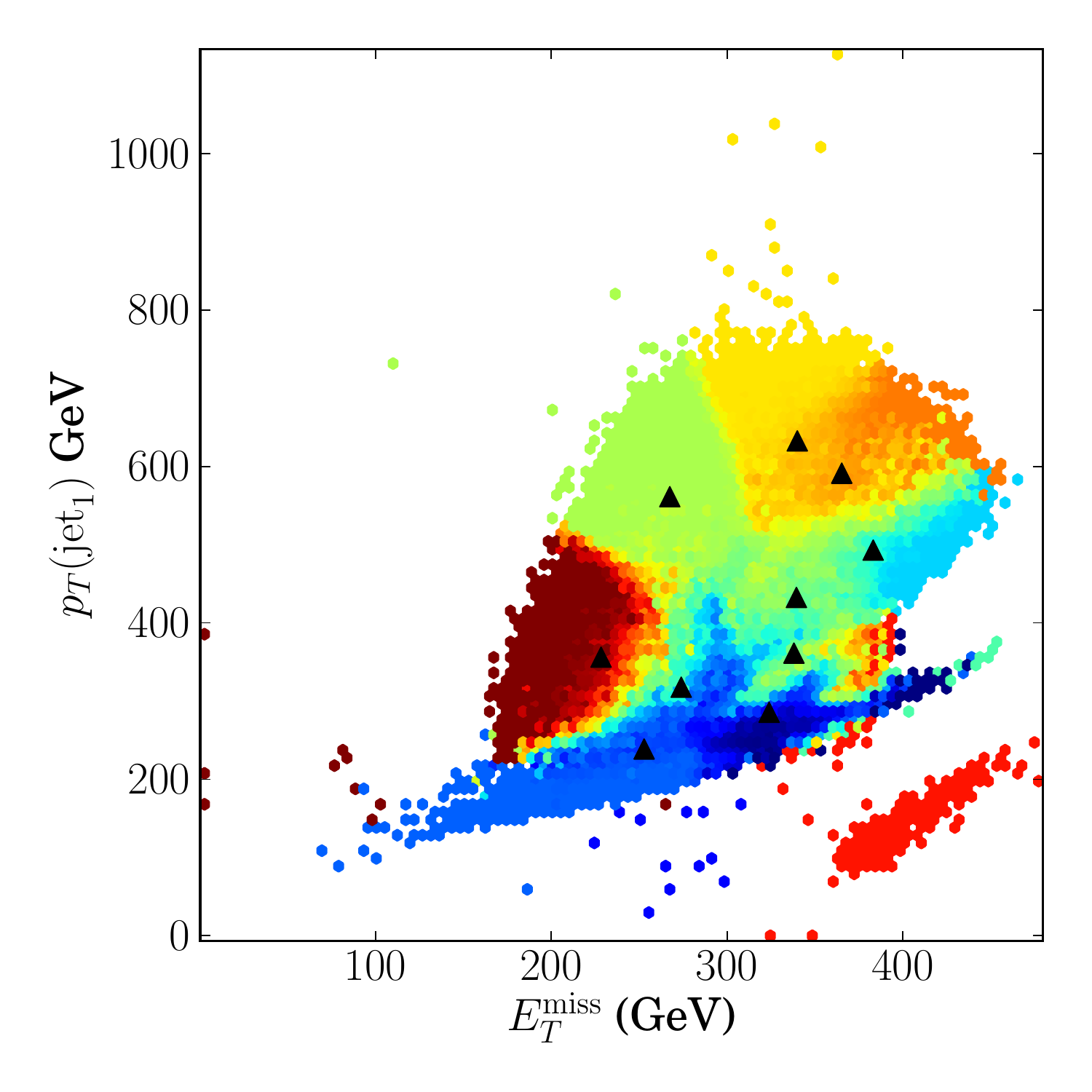}}    
  \end{tabular}}  
  \end{center}
  \caption{Projections of clusters. Colors indicate
    to which cluster a model belongs. Black triangles indicate locations  clusters centroids.} 
  \label{fig:pheno_clusters}
\end{figure}

Various
graphical projections of the clusters are shown in
figures~\ref{fig:susy_clusters} 
and~\ref{fig:pheno_clusters}.
It is clear from figures~\ref{fig:pheno_clusters} that 
 one finds viable models lying in the tails of clusters, far away from centroid positions. 
These models exhibit either low jet activity in the central part
of the detector, but produce high $p_t$ tau leptons, or have low number of high $p_t$ jets (monojets) and low
momentum taus. We have not investigated these models further yet, but it is clear
that standard LHC SUSY searches  assuming presence of high $p_t$ jets for triggering purpose
might fail for such models. 

We have investigated decay branching fractions of the lightest Higgs boson to
$\gamma \gamma$, $ZZ$, $WW$, $\tau \tau$ and $\mu \mu$ for
the ten reference points, compared to these 
of the SM Higgs of the same mass. These branching fractions are typically somewhat higher, alas they do not differ by more than 
$5\%$ from the SM values.

\begin{table}[htbp]
  \caption{Phenomenological parameters of clusters found. The first two columns are the cluster index, id,
    (matches id in table~\ref{tab:clus_susy}) and number of model-points in the
    cluster, n. For each parameter the cluster centroid value, cent, is
    listed along with the minimum, min, and maximum, max, for the
    cluster. Centroid values can be regarded as reference values characterizing given experimental
phenomenology.}   
  \label{tab:clus_pheno}
  \begin{center}
    {\footnotesize
      \begin{tabular}{|p{0.1cm}p{0.65cm}|| p{0.65cm}p{0.65cm}p{0.65cm}|
p{0.45cm}p{0.45cm}p{0.45cm}| p{0.65cm}p{0.65cm}p{0.65cm}|
p{0.45cm}p{0.45cm}p{0.45cm} |p{0.65cm}p{0.65cm}p{0.65cm}|} 
\hline
\multirow{2}{*}{id} & \multirow{2}{*}{n} &
\multicolumn{3}{c|}{$\slashed E_T$ [\GeV]} &
\multicolumn{3}{c|}{$n_\mathrm{jet}$} &
\multicolumn{3}{c|}{$\text{jet}_1 (\pt)$ [\GeV]} &   
\multicolumn{3}{c|}{$n_\tau$} & \multicolumn{3}{c|}{$\tau_1 (\pt)$
[\GeV]}\\
 & & min & cent & max & min & cent & max & min & cent & max & min & cent & max & min & cent & max \\\hline\hline
1 & 16811 & 266.7 & \textbf{323.9} & 439.9&1.5 & \textbf{2.7} & 3.8&206.4 & \textbf{285.7} & 487.4&0.1 & \textbf{0.3} & 0.5&48.3 & \textbf{164.1} & 274.2\\\hline
2 & 6905 & 167.1 & \textbf{273.9} & 332.5&0.1 & \textbf{1.7} & 2.8&31.1 & \textbf{317.8} & 487.8&0.2 & \textbf{0.5} & 0.8&121.8 & \textbf{225.9} & 303.2\\\hline
3 & 6830 & 67.3 & \textbf{252.8} & 316.0&1.1 & \textbf{3.1} & 4.4&56.9 & \textbf{238.8} & 344.1&0.1 & \textbf{0.3} & 0.6&55.2 & \textbf{152.9} & 239.8\\\hline
4 & 11881 & 290.6 & \textbf{383.1} & 467.2&0.9 & \textbf{2.1} & 3.1&316.1 & \textbf{493.1} & 666.5&0.2 & \textbf{0.6} & 1.0&66.5 & \textbf{132.5} & 208.6\\\hline
5 & 9786 & 285.7 & \textbf{339.4} & 451.8&0.8 & \textbf{1.7} & 3.0&243.9 & \textbf{432.5} & 556.4&0.2 & \textbf{0.4} & 0.7&146.2 & \textbf{225.0} & 316.5\\\hline
6 & 10255 & 111.0 & \textbf{267.4} & 331.1&0.0 & \textbf{0.5} & 1.5&376.5 & \textbf{561.4} & 821.5&0.5 & \textbf{0.6} & 1.5&131.9 & \textbf{266.1} & 367.1\\\hline
7 & 11653 & 279.1 & \textbf{339.9} & 389.1&0.0 & \textbf{0.5} & 1.2&473.8 & \textbf{632.7} & 1127.0&0.5 & \textbf{0.6} & 1.1&191.9 & \textbf{280.9} & 367.3\\\hline
8 & 10744 & 300.9 & \textbf{365.2} & 456.6&0.7 & \textbf{1.2} & 2.0&446.2 & \textbf{591.4} & 728.2&0.2 & \textbf{0.5} & 0.8&121.1 & \textbf{234.7} & 298.5\\\hline
9 & 8999 & 259.9 & \textbf{338.0} & 477.2&0.0 & \textbf{0.5} & 2.5&0.0 & \textbf{361.4} & 534.5&0.5 & \textbf{0.6} & 1.2&221.9 & \textbf{307.7} & 476.2\\\hline
10 & 11345 & 0.4 & \textbf{228.3} & 302.6&0.0 & \textbf{0.4} & 2.7&151.3 & \textbf{356.2} & 577.5&0.5 & \textbf{0.7} & 2.0&56.4 & \textbf{273.8} & 385.6\\\hline
\end{tabular}
    }
  \end{center}
\end{table}

\begin{table}[htbp]
  \caption{CMSSM parameters, relict density, 8~TeV CMS LHC production $LO$ cross-section, total NLO k-factors
and the number of expected events with $\tau$ leptons for the centroids of  clusters.  
All models have sign$\mu>0$
    and $m_\mathrm{top} = 173$ \GeV.}
  \label{tab:clus_susy}
  \begin{center}
    {\footnotesize
      \begin{tabular}{|c||c|c|c|c||c|c|c|c|c|c|}\hline
\textbf{id} & $m_0$ [\GeV] & $m_{1/2}$ [\GeV] & $A_0$ [\GeV] &
$\tan \beta$ & $\left<N_\tau\right>_{LO}$&$\left<N_\tau\right>_{NLO}$ & $\ln P$ & $\Omega h^2$ & $\sigma_{LO}$  [fb]& $k_{NLO}$\\ \hline \hline
1 & 821.7 & 937.4 & -2995.0 & 28.4 & 3.6 &4.9 & -1.0 & 0.1 & 0.8 & 1.33\\\hline
2 & 1150.0 & 854.5 & -3318.0 & 37.2 & 19.7& 28.2 & -1.5 & 0.1 &  2.0 & 1.43 \\\hline
3 & 678.4 & 747.7 & -2807.0 & 25.7 & 76.5 &125.5 & -1.9 & 0.1 &  14.1 & 1.64 \\\hline
4 & 278.0 & 740.4 & -1974.0 & 12.3 & 64.1 &85.9 & -1.3 & 0.1 &  7.1 & 1.34\\\hline
5 & 669.4 & 814.6 & -2685.0 & 25.8 & 29.9 &43.1 & -1.1 & 0.1 &  3.5 &  1.44\\\hline
6 & 1045.0 & 961.4 & -2774.0 & 37.9 & 7.5 &8.7 & -1.2 & 0.1 &  0.6 & 1.16 \\\hline
7 & 749.6 & 986.2 & -2450.0 & 29.7 & 6.6 &7.6 & -1.1 & 0.1 &  0.5 & 1.15\\\hline
8 & 481.5 & 824.0 & -2149.0 & 21.5 & 26.4 &33.5 & -1.1 & 0.1 &  2.6 & 1.27\\\hline
9 & 1483.0 & 1070.0 & -3806.0 & 41.0 & 3.5 &4.1 & -1.1 & 0.1 &  0.3 & 1.17\\\hline
10 & 1813.0 & 985.9 & -4153.0 & 46.5 & 6.6 &7.7 & -1.5 & 0.1 &  0.5 & 1.17\\\hline
\end{tabular}}
  \end{center}
\end{table}

\begin{table}[htbp]
  \caption{Higgs and sparticles  masses for the centroids of clusters.}
  \label{tab:clus_spart}
  \begin{center}
    {\footnotesize
      \begin{tabular}{|c||c|c|c|c|c|}\hline
\textbf{id} & $m_{h_0}$ [\GeV] &  $m_{\tilde{t}_1}$ [\GeV]  & $m_{\tilde{g}}$ [\GeV]& $m_{\chi^0_1}$ [\GeV]& $m_{\tilde{\tau}_1}$ [\GeV]\\ \hline \hline
 1 & 125.3 & 997.4 & 2082 & 404.4 & 406.9 \\\hline
 2 & 125.3 & 874.2 & 1932 & 369.7 & 375.8 \\\hline
 3 &  124 & 628.5 & 1690 & 319.6 & 323.5 \\\hline
 4 & 123.7 & 828.3 & 1658 & 313.5 & 314.6 \\\hline
 5 &  125 & 824.3 & 1827 & 348.8 &  354 \\\hline
 6 & 124.5 & 1201 & 2143 &  416 & 421.5 \\\hline
 7 & 124.3 & 1229 & 2177 & 424.9 & 425.3 \\\hline
 8 & 124.2 & 969.2 & 1837 & 351.6 & 353.1 \\\hline
 9 & 125.6 & 1246 & 2384 & 467.9 &  469 \\\hline
 10 & 125.6 & 1185 & 2233 & 432.8 & 437.8 \\\hline
 \end{tabular}
    }
  \end{center}
\end{table}


\section{Conclusions} 
\label{sec:Conclusions}

This work presents a new method for finding and classifying SUSY models that can be potentially discovered in an accelerator experiment, here LHC experiments. 
The method uses an adaptive MCMC algorithm to find interesting models and uses a clustering algorithm to classify the models according to phenomenology. The likelihood map is constructed using an extendible tool chain that incorporates recent limits from multiple sources through the SLHA interface. As the method employs the SLHA interface to communicate between the different tools, it is easily be extendible to other parameter spaces and experimental signatures. For example one can look for interesting regions of GMSB for a two lepton analyses. This amounts to creating a steering file and specify the model, the parameter range of interest, the event topologies counted by \pythia~ and what constraints to take into account, f.ex count \pythia~ events containing $\ell={e,\mu}$, and constrain on $\left\langle N_{2\ell}\right\rangle$. At the moment the actual code is only restricted by limitations to external software tools such as isasugra (allowed parameter spaces) and DarkSUSY (regions with neutralino LSP). In addition the code includes options for doing gridded or uniform random scans in addition to MCMC based methods. The plan is to make the code publicly available in the near future including more parameter spaces and different LSP.

As an example and test of the method the highly constrained CMSSM parameter space has been searched for models that could potentially be discovered using 2012 LHC with  $\tau$-lepton based signatures.

Although simplified models like CMSSM are severely constrained, we are still able to find regions fulfilling recent LHC bounds on Higgs mass and rare $B$-meson decays, and giving relic
density in agreement with WMAP results. All models we found have Higgs BR very close to those of SM. Thus if two sigma excess of Higgs to gamma gamma BR appears to be real then CMSSM is clearly disfavored as observed by \cite{Cao:2012yn}. Ten reference (bench-mark) points exhibiting different
phenomenologies  were found with use of  a {\emph{$g$-means}} clustering algorithm. These reference points can be used to optimize searches. This makes the method very attractive from an experimental point of view, and applying the method to other models and different signatures would be a natural extension of this work. In fact the region found in this search is already part of an effort within the ATLAS astroparticle forum to provide additional model grids taking into account constraints from astro- and astroparticle physics.

The method has proven successful in finding and classifying SUSY models, but could still benefit from several extensions and improvements. More constraints could be added to the likelihoods and more advanced statistical analysis of the simulated data could be incorporated. Another interesting prospect would be to include detector simulations using PGS~\cite{Conway} or DELPHES~\cite{Ovyn:2009tx} to get  
somewhat more realistic estimates for the expected signal, although we are skeptical
about realism of such simulations.  The MCMC algorithm constructed could still benefit from improvements to increase stability and efficiency, in addition to rigorous numerical testing.
Finally a better distance measure for the clustering could allow for precise predictions of expected discovery potential. 

On the more experimental side we find some viable models lying in the tails of clusters formed based
of phenomenological observables. These models exhibit either low jet activity in the central part
of the detector, but produce high $p_t$ tau leptons, or have low number of high $p_t$ jets (monojets) and low
momentum taus. We have not investigated these models further yet, but it is clear
that standard LHC SUSY searches  assuming presence of high $p_t$ jets for triggering purpose
might fail for such models.

\section*{Acknowledgments}

This work has been performed in the scope of the Centre for Dark
Matter Research (DAMARA) at the Department of Physics and
Technology, University of Bergen, Norway. It was
funded by the
Bergen Research Foundation and the University of Bergen, as well as
the Norwegian Research Council in the
framework of  High Energy Particle Physics project.

We would like to give special thanks Therese Sjursen for fruitful 
discussions leading to the start-up of this work and Per Osland for reading through the paper and providing useful comments. We would also like to thank members of the group for Subatomic Physics at University of Bergen for 
their support.

\begin{appendices}
\section{Scan Implementation, clustering and cross-checks}
\label{sec:appendix}

The implementation of the scan was written in \python. It was capable of
running multiple MCMC chains in parallel. The scan was initiated using a random
sample containing roughly 100 points to find an estimate for the proposal
distribution. The initial points were required to pass all discrete constraints,
be within $2\sigma$ of the best fit values for $\textbf{Br}(b\rightarrow s+\gamma)$, $\textbf{Br}\left(B_s\rightarrow \mu\mu\right)$, $m_{h0}$, while having $\Omega h^2$ of the right order of magnitude and an expected number of produced $\tau$ leptons, $\left\langle N_\tau\right\rangle>1$. The values are summarized in table~\ref{tab:Icons}, together with the distributions used for initializing the proposal distribution. 
 
\renewcommand*\arraystretch{1.2} 
\begin{table}[htbp]
  \caption{Experimental constraints used in the  initial sampling.}
  \label{tab:Icons}
  \begin{center}    
      \begin{tabular}{|c|c|c|}\hline
        \textbf{Constraints}&\textbf{Range}&\textbf{Distribution}\\\hline\hline
        $\Omega h^2$& $\left(0, 0.2\right]$&Flat\\   
        $\textbf{Br}\left(b\rightarrow s+\gamma\right)$ & $\left[2.71,4.39\right]\cdot 10^{-4}$&Gaussian\\
        $\textbf{Br}\left(B_s\rightarrow \mu\mu\right)$ & $\left[0.2,6.2\right]\cdot 10^{-9}$&Gaussian\\  
        $m_{h0}$ &$ \left[122.3,128.7\right]\ \GeV $&Gaussian\\
        $\left\langle N_\tau\right\rangle$&$\left[1,\infty\right)$&Flat\\\hline
      \end{tabular}
  \end{center} 
\end{table}   

Five clusters were established from the initial parameter space points with the \emph{$k$-means} algorithm. This number was found to be sufficient to give a reasonable approximation of the likelihood  distribution of the sample.
From the initial sample, ten chains were initiated with two chains
starting from each cluster. Before sampling started, each chain was
required to reach a minimum likelihood to be included in the
sample. This was chosen to be $2 \sigma$ away from the central value for
 $\textbf{Br}(b\rightarrow s+\gamma)$, $\textbf{Br}\left(B_s\rightarrow \mu\mu\right)$, $m_{h0}$, $\Omega h^2$, in addition to $\left\langle N_\tau\right\rangle\geq1$, corresponding
to the likelihood, $\ln P_{\min}\sim-4\cdot\frac{4}{2}-0.5=-8.5$.
  
The proposal distribution was updated at intervals $\Delta N=1000$ steps by 
adding the new sample points and recalculating cluster means and
covariances. The new points were added without weights since 
they were already a product of weighted sampling. For practical purposes
we end the optimization after 10 000 steps, which was found to be sufficient to give a good proposal estimate. By fixing the proposal after a certain number of steps the algorithm also satisy the neccessary conditions to ensure asymptotic convergence toward the true likelihood distribution, since the algorithm becomes equivalent to running a set of independent Metropolis-Hastings chains. The search chains were run in parallel on twenty cores for roughly $200$ hours, 
resulting in a sample size of $N=2\ 076\ 133$, corresponding to $105\ 209$ unique models. From this 
sample $848$ outliers (corresponding to $66$ unique models) 
with log-likelihood $ln~P<-8.5$ were removed.  

In order to illustrate the effects of the different experimental and theoretical constraints, 
low energy properties were calculated for $300~000$ models sampled uniformly  within the search range. 
The computationally expensive \pythia\ simulations where not done for these models and a looser relic 
density constraint compared to the one used for MCMC initialization was used to get sufficient data to describe 
the qualitative features of the constraint.    

\begin{figure}[htbp]
 \begin{center}
   \makebox[\textwidth]{
   \begin{tabular}{cccc}
   
   \subfloat{\includegraphics[width=35mm]{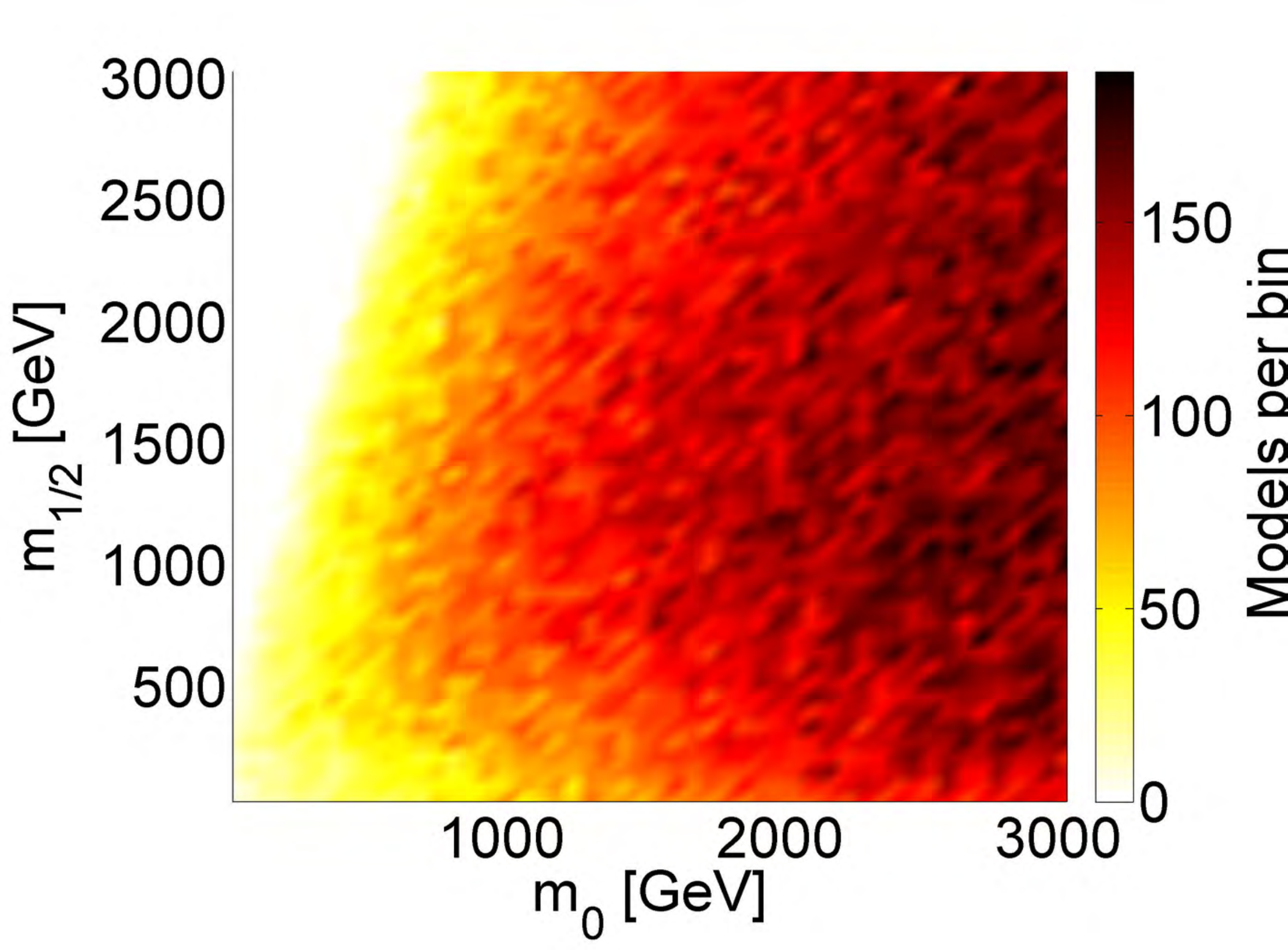}}
   &
   \subfloat{\includegraphics[width=35mm]{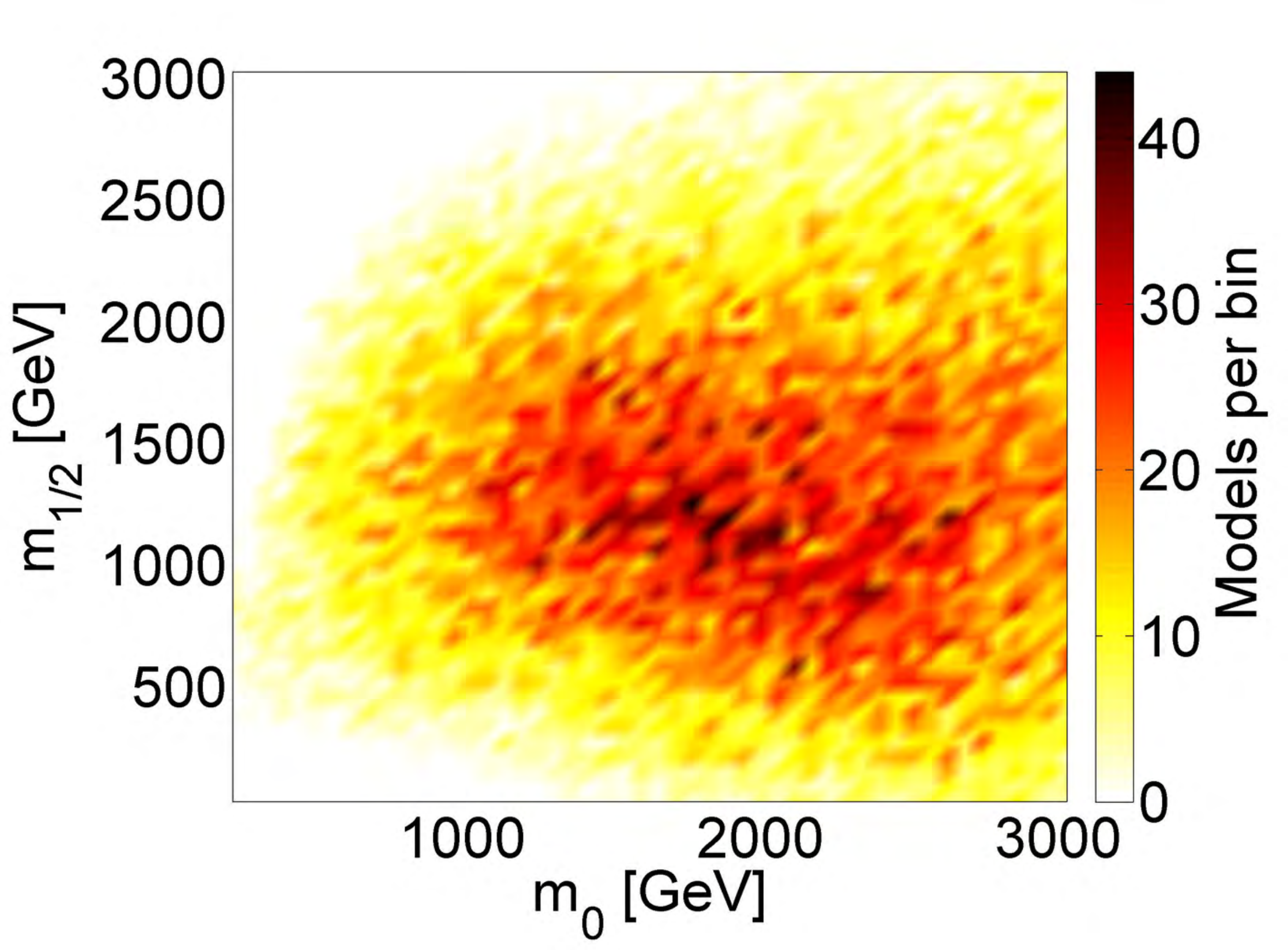}}
   &
   \subfloat{\includegraphics[width=35mm]{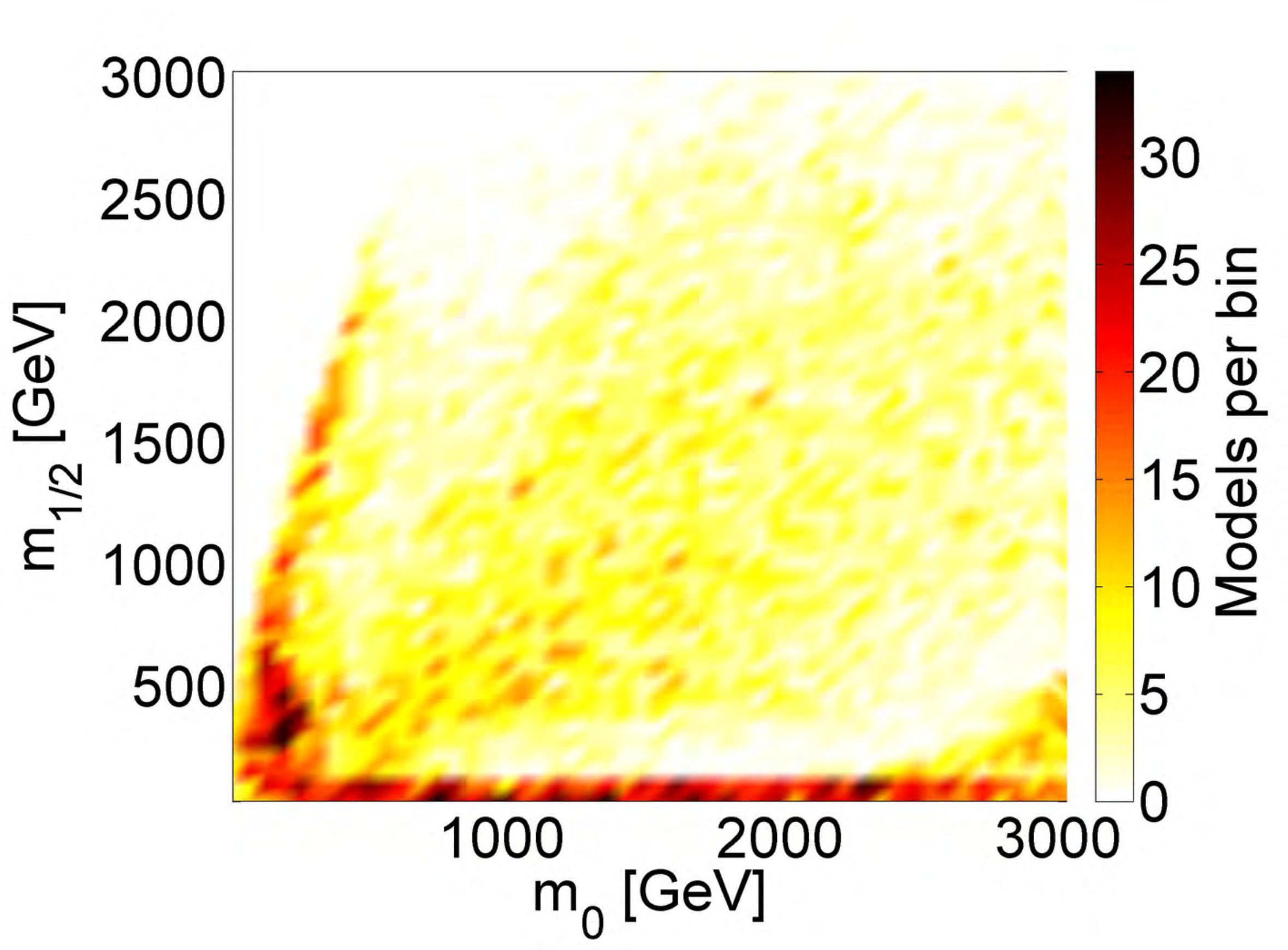}}
   &
   \subfloat{\includegraphics[width=35mm]{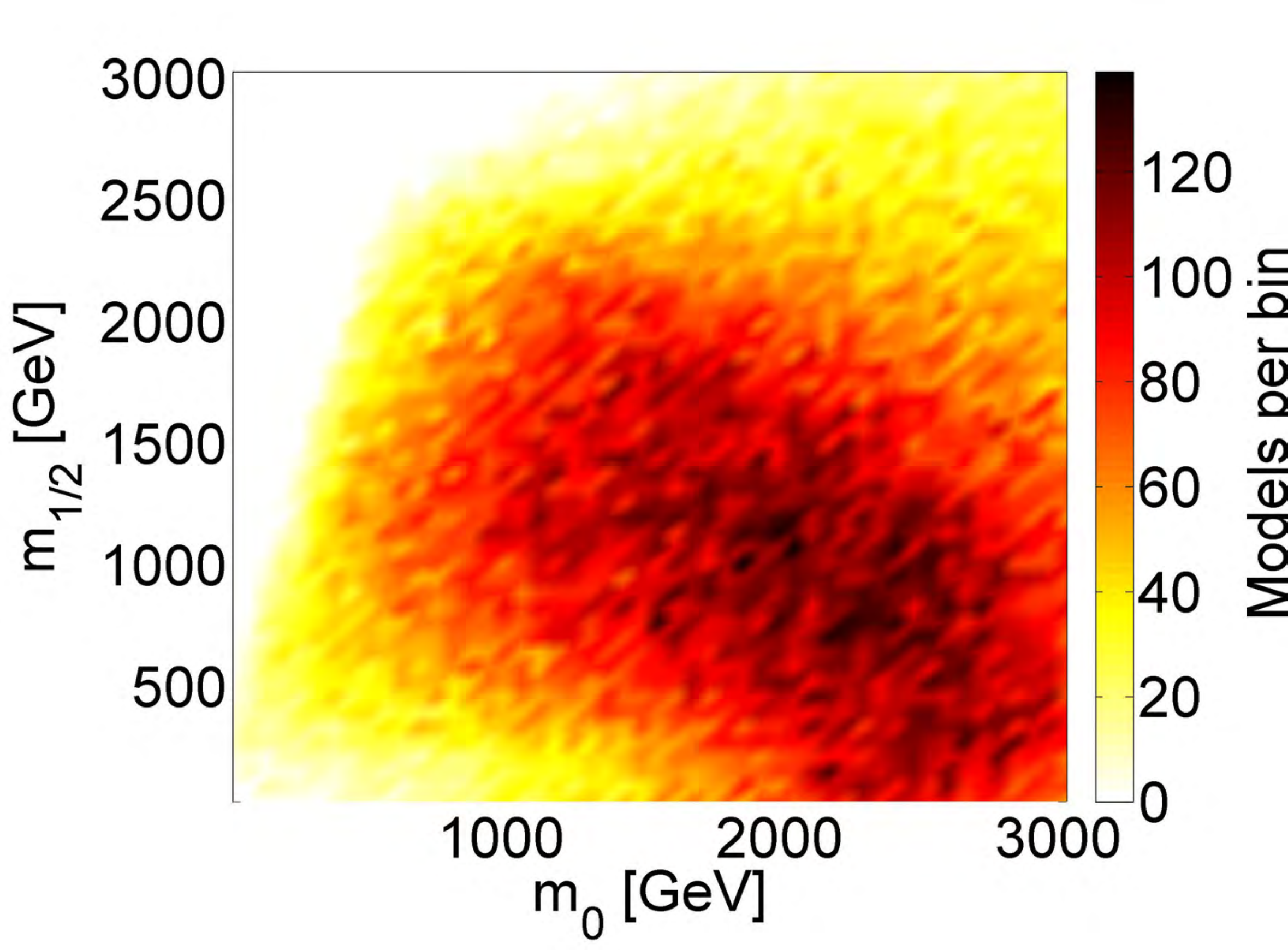}}
   \\
   \subfloat[Theoretical constraints]{\includegraphics[width=35mm]{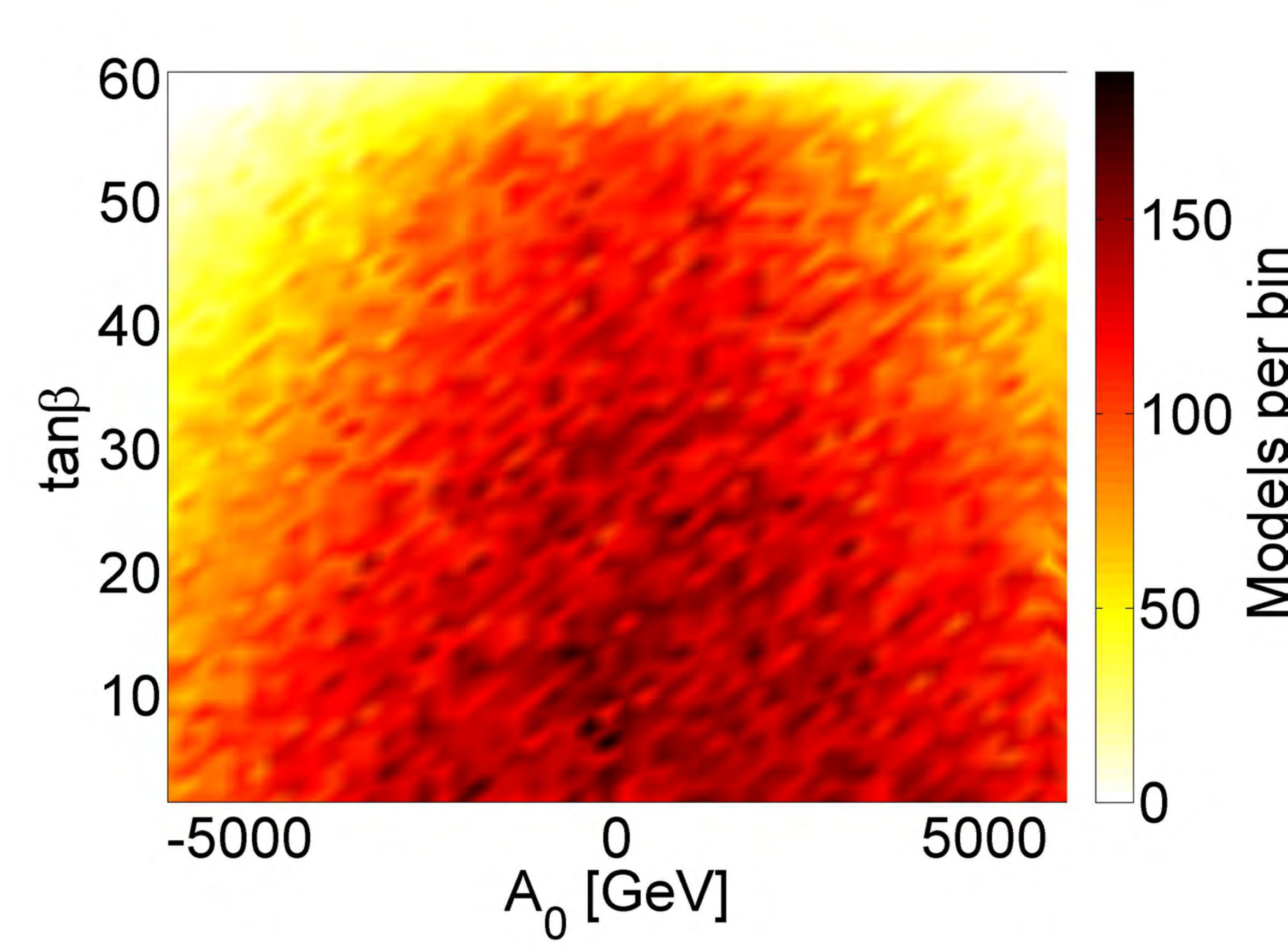}}
   &
   \subfloat[Th.C+$m_{h0}$]{\includegraphics[width=35mm]{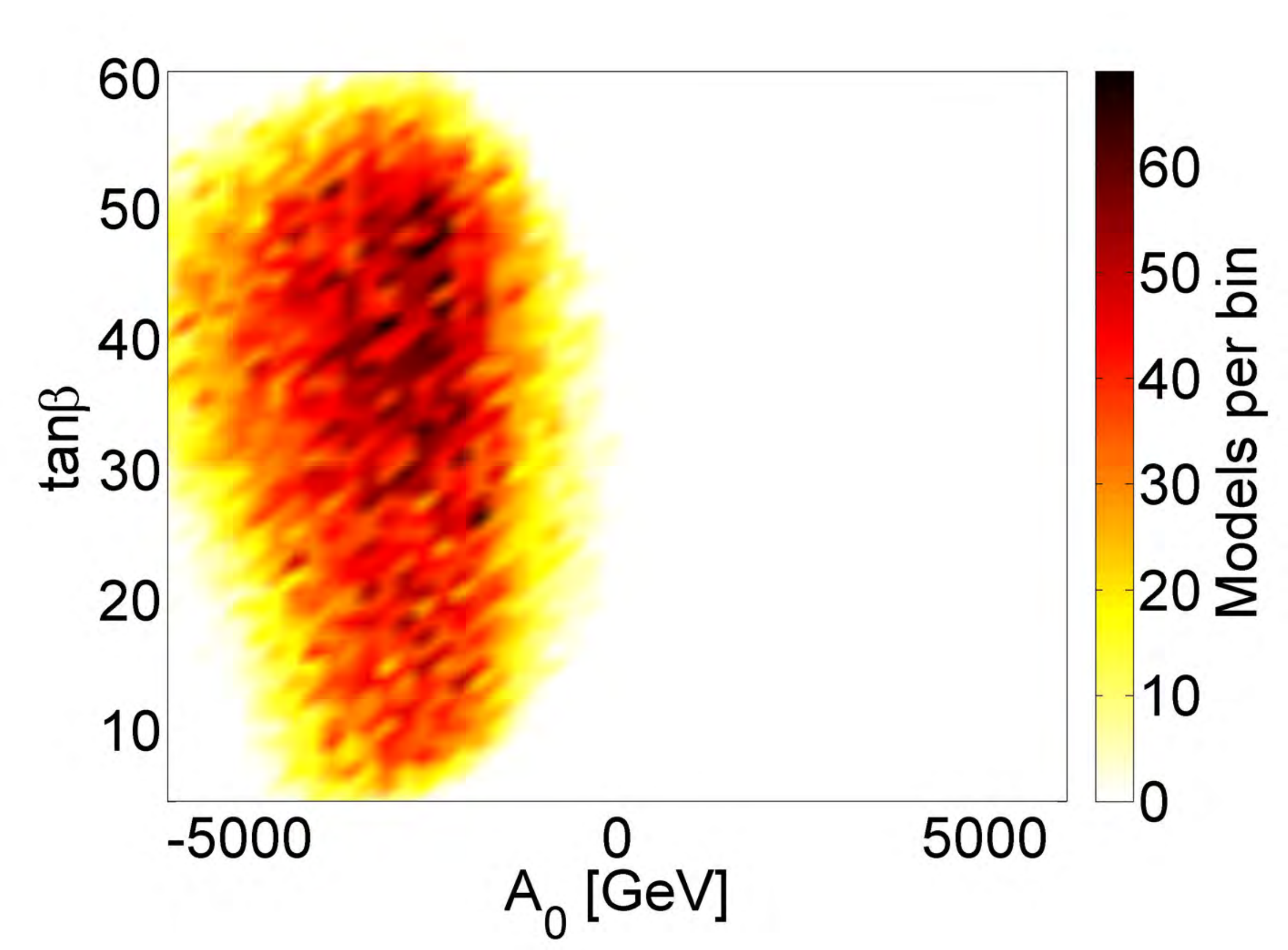}}
   &
   \subfloat[Th.C+$\Omega h^2$]{\includegraphics[width=35mm]{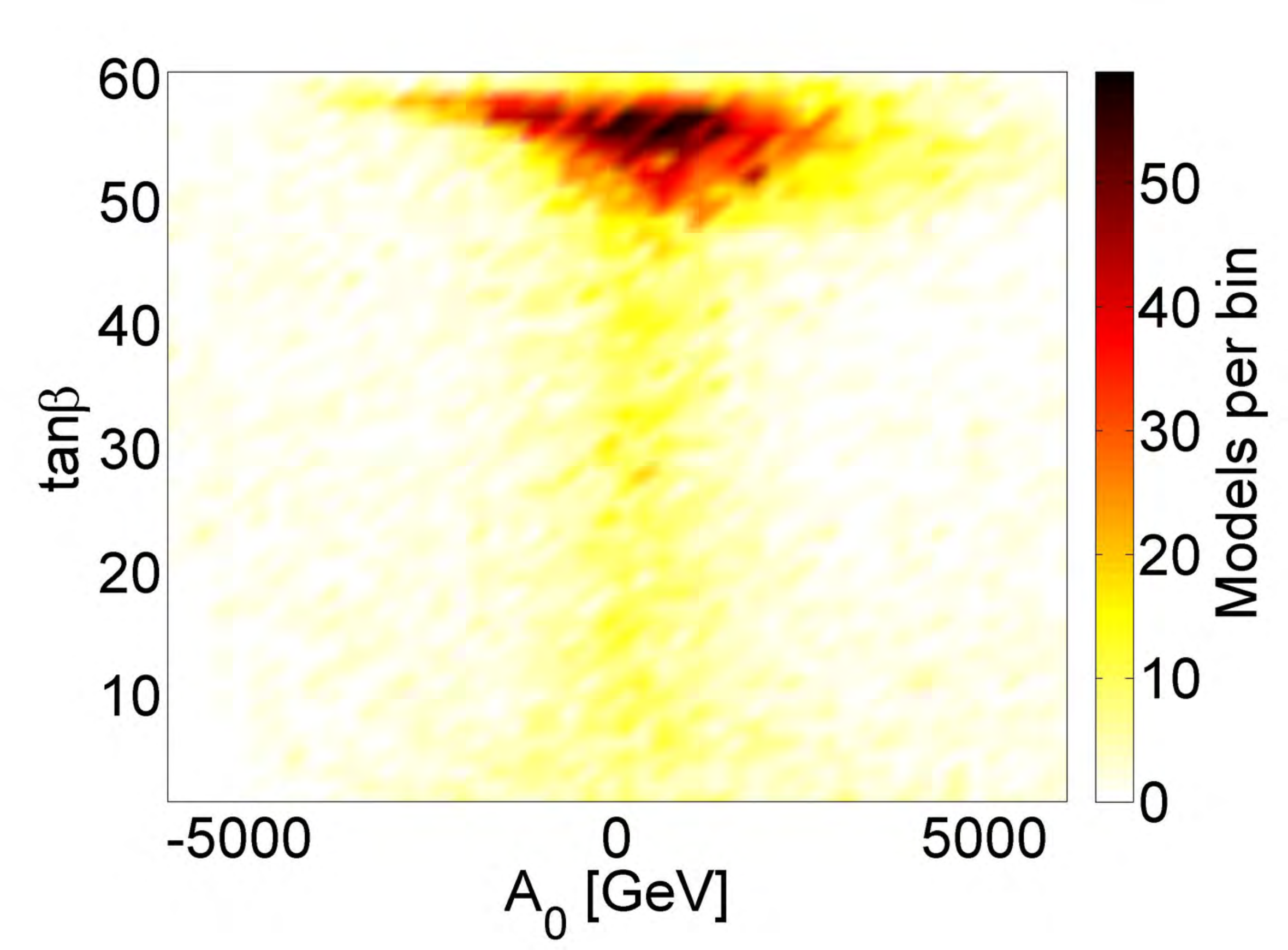}}
	 &
   \subfloat[Th.C+$\textbf{Br}_{B_s\rightarrow\mu\mu,b\rightarrow s\gamma}$]{\includegraphics[width=35mm]{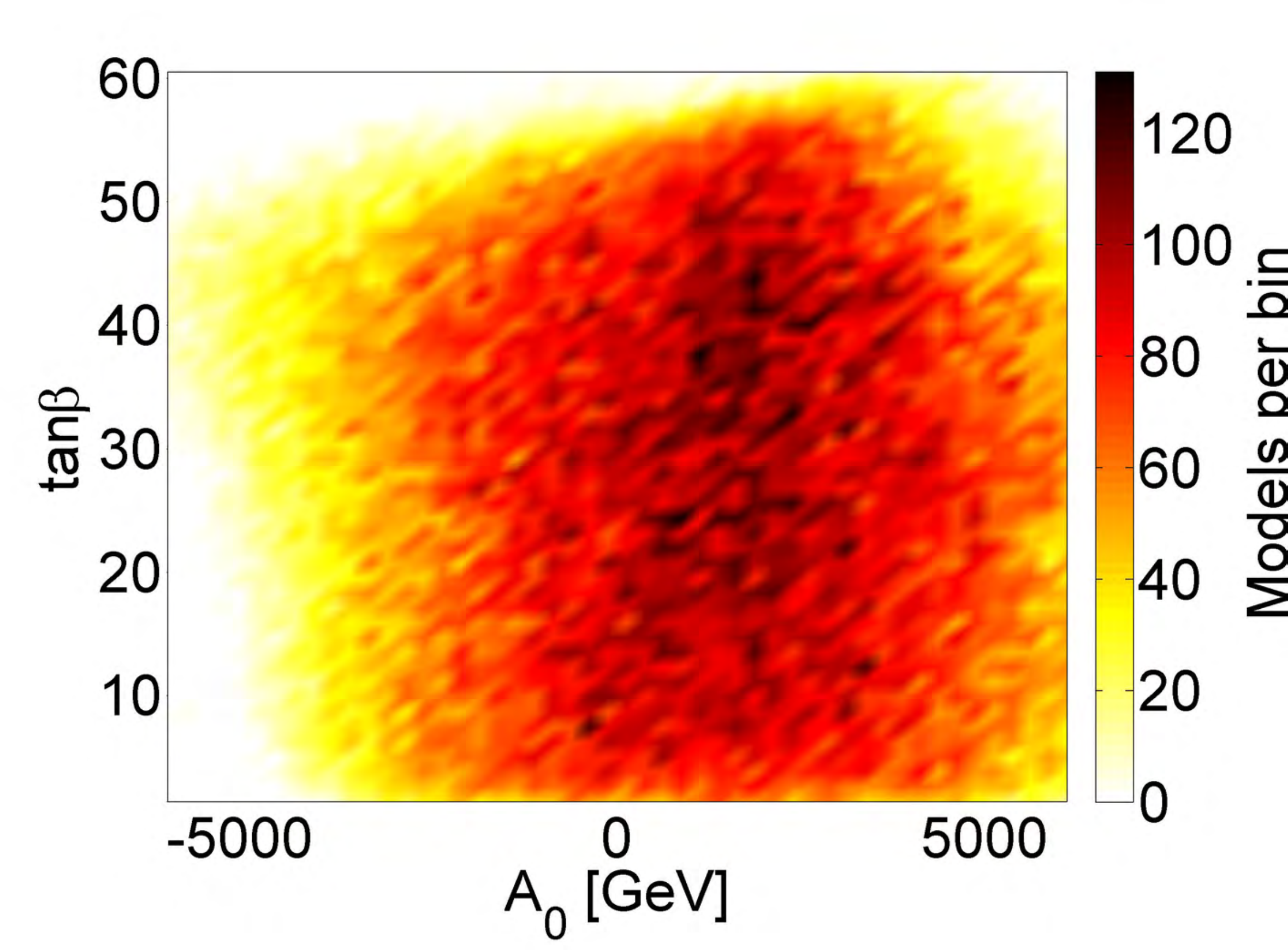}}
   \end{tabular}}
 \end{center}
 \caption{2D-histograms in $m_0,m_{1/2}$ and $A_0,\tan\beta$-planes showing the effects of different constraints. The constraints used corresponds to requirements chosen for the initial sample given in table~\ref{tab:Icons}}
 \label{fig:Unicuts} 
\end{figure}

As a cross-check with the vast literature
on the subject (see for example~\cite{Profumo:2011zj,Allanach:2005kz,Ellis:1999mm})
we briefly describe the effects of the most important constraints by visualizing how 
the initial selections affect the model density. 
The effects are illustrated in figure~\ref{fig:Unicuts}. We observe, in agreement with results of 
\cite{Profumo:2011zj,Fowlie:2012im} 
that:

\begin{itemize}
	\item \textbf{Theoretical constraints}\\
	Theoretical constraints remove  the low $m_0$ and $m_{1/2}$ regions primarily avoiding 
a $\tilde{\tau}_1$-LSP and tachyonic sparticles. The excluded regions becomes larger at large $\tan\beta$ and $|A_0|$,
 and for large values of $A_0$ a considerable part of the low mass regions, 
$m_0,m_{1/2}\lesssim 1000\ \GeV$  gives tachyons.

	\item \textbf{Higgs mass} $\mathbf{m_{h0}\in\left[122,128\right]}$\\
	Requiring a $125.5\ \GeV$ Higgs mass, with positive $\mu$ excludes all positive values of $A_0$ within the selected $m_0,m_{1/2}$-range. 
For large negative values of $A_0$ however, the $\tilde{t}$-loop corrections to the Higgs mass become large.
 This is the main
 reason for the asymmetry in $A_0$ seen in \ref{fig:LhoodUni}.
 At large values of $m_{1/2}$ and $m_{0}$ the \feynhiggs~calculations of the Higgs mass corrections become inaccurate 
and thus we excluded these regions.   
	
	\item \textbf{Relic Density }$\mathbf{\Omega h^2<1}$\\
	As is well known, the relic density Dark Matter in CMSSM is generally 
orders of magnitude larger than allowed by WMAP and PLANCK results 
\cite{Baer:2009bu,Baer:2005jq,Ellis:1999mm}, 
apart from special regions where the relic density is suppressed by  resonant neutralino annihilation 
or co-annihilation cross-sections.  
The low $m_{1/2}$ region where $m_{\tilde{\chi}^0_1}\lesssim10$, the relic density is mainly suppressed 
through $\chi$-annihilation to fermions through sfermion exchange (low $m_0$), and to $W,Z$ pairs 
(high $m_0$). This region is excluded primarily by Higgs mass requirements. 
Along the $\tilde{\chi}^0_1$-LSP boundary the relic density is 
reduced by $\chi-\tilde{\tau}$-coannihilation, since the coannihilation cross-section
is significantly enhanced due to mass degeneracy between the lightest stau and the lightest neutralino. 
The middle region in the mass plane, the  well known  Higgs funnel \cite{Roszkowski:2001sb},
corresponds to high $\tan\beta$ models with $m_{\tilde{\chi}^0_1}\sim 1/2m_{H^0,A^0}$, 
giving an increase in $\chi-\chi$-annihilation through heavy neutral higgs bosons, ($H^0$,$A^0$). 
The preference for $A_0\sim 0$ arises mainly from the fact that large parts of the low $m_{1/2}$ regions 
exhibit charged LSP or tachyonic particles for large values of $|A_0|$, as off-diagonal terms
in the third generation sfermion mass matrices grow with $|A_0|$. 
The preference for high $\tan\beta$ is in  part due to  
the additional relic density suppression through the Higgs channel $\chi$-annihilation.   
	
	\item \textbf{Rare Decays }$\mathbf{\textbf{Br}(B_s\rightarrow\mu\mu)<4.5\cdot10^{-9},\textbf{Br}(b\rightarrow s+\gamma)\in\left[3,4\right]\cdot 10^{-4}}$\\
	Of the constraints on decays, $B_s\rightarrow\mu\mu$ poses the most stringent one, as the SUSY
contribution grows like $\tan\beta^6$.
This branching fraction tends to get too large at low values of $m0$ and $m_{1/2}$.
The size of the excluded area in the mass plane increases with increasing $tan\beta$  and decreasing $|A_0|$. 
$\textbf{Br}(b\rightarrow s+\gamma)$ is generally too low compared to the central
experimental  value of $3.55\cdot 10^{-4}$ and excludes large parts of the low $m_{1/2}\lesssim 500$ range,
stretching as far as $m_0\sim2000$ for high values of $\tan\beta$ and low $|A_0|$. Too high $\textbf{Br}(B_s\rightarrow\mu\mu)$ and too low $\textbf{Br}(b\rightarrow s+\gamma)$, together with the requirement of non-tachyonic sparticles constrains 
the lowest allowed values of $m_{1/2}$.    
	     
\end{itemize}

The properties of selected high likelihood models are presented in the results section,
\ref{sec:Results}.
The relatively wide range of values for SUSY masses and values of $\tan\beta$ for the
selected models  lead to a
 wide range of values of phenomenological properties such as average 
$\slashed E_T$,the average missing energy per SUSY event, $p_T(\tau_1),p_T(\mathrm{jet}_1)$, the average $p_T$ 
of the leading $\tau$ and the leading jet, and $n_{\tau},n_\mathrm{jet}$, the average number of $\tau$'s/jets per 
SUSY event, see figures~\ref{fig:muPhenopT} and~\ref{fig:muPhenon}. 
In order to construct reference models that cover these different phenomenological 
properties the sample was clustered according to the phenomenological observables listed
above.
In order to avoid bias from the scale of the different variables, each variable $x$ is first transformed as $x'=(x_i-\bar{x})/\sigma_x$ so that the mean $\bar{x}'=0$ and variance $\sigma_{x'} ^{2} = 1$.
Because the non-Gaussian nature of clusters  the
\emph{$g$-means} algorithm often fail and split too often. To remedy this
the constraining parameters mentioned in section~\ref{sec:Clustering}
are used. By setting an approximate maximum number of possible
clusters $n_\mathrm{max}$, one gets $\min_P = \lceil N_\mathrm{OK} /
n_\mathrm{max} \rceil$ for the minimum number of points in a cluster and
$\min_s = \lceil \log_2 n_\mathrm{max} \rceil$ for the maximal
splitting depth. The maximal number of iterations per split attempt
was set to $\max_i = 20$. Here the number is chosen to be well above
the final number of clusters but low enough, for this case
$n_\mathrm{max}=100$ was found to be appropriate. The minimal cluster
distance parameter was set to $\min_d=1.3$. The optimization was run
$n_\mathrm{avg}=7$ times and an average of 9.8 clusters were
found. Thus, one of the results with 10 clusters was picked at
random. The properties of models at the centroids of these clusters, which
can be seen as reference models for search optimization, are presented 
in the results section~\ref{sec:Results}.

\end{appendices}

\bibliographystyle{JHEP}
 
\providecommand{\href}[2]{#2}\begingroup\raggedright\endgroup



\end{document}